\documentclass[12pt,a4paper]{iopart}

\usepackage{iopams}  
\expandafter\let\csname equation*\endcsname\relax
\expandafter\let\csname endequation*\endcsname\relax
\usepackage{amsmath}
\usepackage{amssymb}
\usepackage{graphicx}
\usepackage{mathrsfs}
\usepackage{relsize}
\usepackage{appendix}
\usepackage{float}
\usepackage[colorlinks=true,allcolors=blue]{hyperref}
\usepackage{tablefootnote}
\usepackage{footmisc}
\usepackage{soul}
\allowdisplaybreaks[0]
\newcommand{\ud}{\mathrm{d}}
\newcommand{\ord}[2]{\underset{^{(#1)}}{#2}{}}

\DeclareMathOperator{\STF}{STF}

\usepackage{xcolor}
\usepackage[normalem]{ulem}
\let\oldsum\sum
\let\oldprod\prod

\renewcommand{\sum}{\mathlarger{\oldsum}}
\renewcommand{\prod}{\mathlarger{\oldprod}}

\numberwithin{equation}{section}

\begin{document}

\title[]{Higher memory effects and the post-Newtonian calculation of their gravitational-wave signals}

\author{S.\ Siddhant$^1 \footnote{Author to whom any correspondence should be addressed}$, \; Alexander M.\ Grant$^{1, 2}$, \; David A.\ Nichols$^1$ }

\address{$^1$ Department of Physics, University of Virginia, P.O.~Box 400714,
  Charlottesville, Virginia 22904-4714, USA}
\address{$^2$ School of Mathematical Sciences, University of Southampton,
  Southampton, SO17 1BJ, United Kingdom}
\ead{sbp4ab@virginia.edu, a.m.grant@soton.ac.uk, david.nichols@virginia.edu}
\vspace{10pt}
\begin{indented}
\item[]\today
\end{indented}

\begin{abstract}
A new hierarchy of lasting gravitational-wave effects (the higher memory effects) was recently identified in asymptotically flat spacetimes, with the better-known displacement, spin, and center-of-mass memory effects included as the lowest two orders in the set of these effects.
These gravitational-wave observables are determined by a set of temporal moments of the news tensor, which describes gravitational radiation from an isolated source.
The moments of the news can be expressed in terms of changes in charge-like expressions and integrals over retarded time of flux-like terms, some of which vanish in the absence of radiation.
In this paper, we compute expressions for the flux-like contributions to the moments of the news in terms of a set of multipoles that characterize the gravitational-wave strain.
We also identify a part of the strain that gives rise to these moments of the news.
In the context of post-Newtonian theory, we show that the strain related to the moments of the news is responsible for the many nonlinear, instantaneous terms and ``memory'' terms that appear in the post-Newtonian expressions for the radiative multipole moments of the strain.
We also apply our results to compute the leading post-Newtonian expressions for the moments of the news and the corresponding strains that are generated during the inspiral of compact binary sources.
These results provide a new viewpoint on the waveforms computed from the multipolar post-Minkowski formalism, and they could be used to assess the detection prospects of this new class of higher memory effects.
\end{abstract}

\section{Introduction}\label{sec:Introduction}

In 1962, Bondi \textit{et al}.~\cite{Bondi:1962px} published a study of the solutions of Einstein's equations far from an isolated and radiating axisymmetric source, which were written in a set of well adapted coordinates (Bondi coordinates). 
Specifically, outgoing Bondi coordinates are based on a retarded time variable $u$, an areal radius $r$, and angles on a unit two-sphere.
The work of~\cite{Bondi:1962px} was generalized by Sachs~\cite{Sachs:1962zza} to spacetimes without axisymmetry soon thereafter.
Several novel features of these asymptotically flat spacetimes were revealed by these works~\cite{Bondi:1962px,Sachs:1962zza}:
\begin{enumerate}

\item Einstein's equations in these coordinates can be solved hierarchically with initial data on a null hypersurface of constant $u$ and boundary conditions imposed on the metric as $r\rightarrow\infty$.

\item The solutions of these equations demonstrated there is a \emph{news} tensor with two independent components that characterizes the gravitational radiation emitted from the isolated source and causes the mass of the source to decrease when it is non-zero.

\item The asymptotic symmetry group that preserves the leading-order metric in an expansion in $1/r$ and the Bondi gauge conditions was determined to be an infinite-dimensional generalization of the ten-dimensional Poincar\'e group known as the Bondi-Metzner-Sachs (BMS) group.

\end{enumerate}
Much like the Poincar\'e group, the BMS group admits Lorentz transformations, but the BMS group also allows for ``angle-dependent translations'' called ``supertranslations'' that include the four-dimensional spacetime translations of the Poincar\'e group as a subgroup. 
It is the supertranslations that make the BMS group infinite dimensional. 

The BMS formalism has played an important role more recently in studying another facet of gravitational radiation, the \emph{gravitational-wave memory effect} (also referred to as the \emph{displacement memory effect}).
The memory effect is a lasting gravitational-wave strain that persists after the waves pass by an observer far from an isolated source, and it produces a(n in principle) measurable relative displacement of initially comoving, freely falling test masses.
It was originally computed in contexts other than the BMS formalism and different astrophysical sources of the memory effect were identified over the course of several decades.
The so-called linear (or ``ordinary''~\cite{Bieri:2013ada}) part of the effect was first computed by Zeldovich and Polnarev in 1974~\cite{Zeldovich:1974gvh}, although the term ``memory effect''  was only introduced later by Braginsky and Grishchuk~\cite{Braginsky:1985vlg}. 
The nonlinear (or ``null'') portion of the effect was calculated by Christodoulou~\cite{Christodoulou:1991cr} and in the context of the multipolar post-Minkowski formalism by Blanchet and Damour~\cite{Blanchet:1992br}. 

In the BMS formalism, the mathematical expressions that give rise to the linear and nonlinear memory effects appear clearly in one of the asymptotic Einstein equations (the same one that was used by Bondi \textit{et al}.~\cite{Bondi:1962px} to compute the well-known mass-loss formula).
Moreover, the BMS formalism allows for the memory effect to be interpreted in terms of the BMS supertranslation symmetries and the corresponding charges conjugate to these symmetries, the supermomenta~\cite{Geroch:1977jn,Wald:1999wa}.
Specifically, the component of Einstein's equation from which the memory can be computed contains equivalent information to the flux-balance laws for supermomentum~\cite{Strominger:2014pwa,Ashtekar:2014zsa}.
The existence of the memory effect can then be understood, in this context, as a consequence of the supermomentum balance laws.
Moreover, the lasting strain associated with the memory effect exists because the initial and final ``canonical'' rest frames before and after the waves differ by a supertranslation~\cite{Strominger:2014pwa,Ashtekar:2014zsa,Flanagan:2015pxa}. 

The BMS formalism has also allowed other types of memory effects to be identified.
Pasterski \textit{et al}.~\cite{Pasterski:2015tva} deduced the existence of a new memory effect related the fluxes of angular momentum from the source (see also~\cite{Flanagan:2015pxa,Nichols:2017rqr}), which was called ``spin memory.''
A closely related memory effect corresponding to the flux of the electric-parity part of angular momentum, called ``center-of-mass (CM) memory,'' was identified not long after~\cite{Nichols:2018qac}. 
Both of these memory effects could also be interpreted in terms of flux-balance laws of extensions of the BMS group: in particular, the extended BMS group proposed by Barnich and Troessaert~\cite{Barnich:2009se, Barnich:2010eb} or the generalized BMS group of Campiglia and Laddha~\cite{Campiglia:2014yka,Campiglia:2015yka}. 
Unlike the displacement memory effect associated with supermomentum fluxes, the spin and CM memories were related to lasting changes in the time integral of the shear (which produces part of the lasting displacement of freely falling test masses when the test masses are no longer initially comoving~\cite{Flanagan:2018yzh}).
In this case, however, there is not a corresponding interpretation of the memory effects as arising from the initial and final canonical rest frames of the system differing by a generalized or extended BMS transformation~\cite{Compere:2018ylh}.

While the three memory effects (displacement, CM and spin) have close connections with the symmetries and conserved quantities of the BMS framework and its extensions and generalizations, there are a large class of lasting gravitational-wave phenomena, called ``persistent observables'' by~\cite{Flanagan:2018yzh}, which did not need to be computed in the context of asymptotically flat spacetimes and did not have obvious relationships to symmetries and conserved quantities.
One of the observables of~\cite{Flanagan:2018yzh}, the curve deviation, 
is a generalization of the (in-principle) procedure used to measure the memory effect.
Specifically, it allows the test masses that are used to measure the relative separation to be in relative acceleration rather than freely falling, and they are no longer assumed to be initially comoving. 
The relative separation will change because of both the initial relative velocity and the relative acceleration, but once these kinematical effects are removed, there is a residual change in the final relative separation that depends on the spacetime curvature and its time integrals, as well as the initial separation, velocity, acceleration, and all initial time derivatives of the acceleration. 
This residual change was what was described as the curve deviation observable.

The curve deviation was then computed in asymptotically flat spacetimes in the BMS framework in~\cite{Grant:2021hga}. 
In this context, the leading-order expression in $1/r$ for the curve deviation could be be written in terms of a set of temporal \emph{moments of the news tensor}, which are closely related to the Mellin transforms of the news computed in~\cite{Compere:2022zdz}.
The zeroth moment of the news is simply the displacement memory; the first moment of the news is related to the ``drift'' (or, in the older terminology of~\cite{Flanagan:2018yzh}, ``subleading displacement'') memory, which has the spin and CM memories as its magnetic and electric parts on the sphere; and the second moment of the news is related to what was named the ``ballistic'' memory in~\cite{Grant:2023ged}.
It was demonstrated in~\cite{Grant:2021hga} that these moments of the news could be expressed in terms of changes in charge-like quantities (expressions that did not vanish when there was no radiation) and retarded-time integrals of fluxes (where the integrand vanishes in the absence of radiation).
The work of \cite{Compere:2022zdz} and~\cite{Geiller:2024bgf} considered the connection between these charge-like quantities and charges that form a representation of the loop $w_{1+\infty}$ algebra that appears in the celestial holography program (see~\cite{Donnay:2023mrd} for a review) and in the context of the Weyl-BMS extension of the BMS group~\cite{Freidel:2021fxf,Freidel:2021cjp,Freidel:2021ytz}.
The curve deviation as computed in~\cite{Grant:2021hga} then may ultimately be constructed from quantities that arise from flux-balance laws that are related to symmetries. 

In this paper, however, we will put a greater emphasis on computations of these moments of the news in terms of multipoles of the gravitational-wave strain and in the context of an approximation method that allows these moments to be determined for astrophysical sources of gravitational waves.
We also compute time-dependent gravitational-wave signals that are related to these moments of the news, which are quantities that can be measured more straightforwardly by existing gravitational-wave detectors, such as LIGO, Virgo and KAGRA. 
To do so, we first define slightly different moments of the news from those used in~\cite{Grant:2021hga}; these new moments are just repeated integrals of the news over retarded time and have a simpler connection to the corresponding gravitational-wave signals.
These moments of the news and gravitational-wave signals are what we refer to as ``\emph{higher memory effects},'' the phrase that appears in the title of this paper.

These moments of the news have flux and charge contributions, and we will focus on the flux contributions in this paper.
For higher moments of the news, the flux terms involve an increasing number of integrals of fluxes that have appeared in lower moments of the news; throughout this paper, we will focus on the ``new'' parts of the fluxes, which do not appear in the expressions for lower moments of the news.
These fluxes can be written in terms of products of the shear and news (and their angular derivatives), and we derive general multipolar expressions for these new fluxes in terms of a set of radiative multipole moments of the shear.
Starting at the second moment of the news, we also compute the multipolar expansion of ``pseudo-fluxes'' (quantities that depend on the shear or other so-called ``non-radiative'' Bondi metric functions).
We compute the fluxes and pseudo-fluxes necessary to determine the second moment of the news; however, our results could be extended to compute higher moments of the news.

The general multipolar results in this paper are then applied to understand features of the post-Newtonian (PN) expressions for the gravitational waveform for general PN sources.
In particular, we demonstrate that up to a total second time derivative, the instantaneous and memory terms in the 3.5~PN-order expressions for the radiative mass quadrupole and octupole multipole moments as a function of the canonical moments (see~\cite{Blanchet:2013haa}) are equivalent to the gravitational-wave memory signals that arise from the zeroth, first, and second moments of the news.
This provides a new perspective on the relationship between the Bondi-Sachs and multipolar post-Minkowskian framework that has been investigated recently in~\cite{Blanchet:2020ngx,Blanchet:2023pce}.

We also use the multipolar expansions to compute the leading PN expressions for the moments of the news (and their corresponding gravitational-wave signals) that are generated by the leading gravitational-wave strain produced during the inspiral of nonspinning compact objects on quasicircular orbits.
We recover the leading-order non-oscillatory effects for the nonlinear displacement and spin memory effects given in Refs.~\cite{Favata:2008yd} and~\cite{Nichols:2017rqr}, respectively, which are computed from the leading PN oscillatory waveforms without memory.
We also compute ``oscillatory memory'' terms, which are nonlinear contributions to the oscillatory waveform modes of the strain that arise from the same flux and pseudo-flux terms used to compute the more familiar non-oscillatory memory effects.
Our PN calculation of the oscillatory CM memory corrects the result in~\cite{Nichols:2018qac}; however, we also complete calculations of the non-oscillatory and oscillatory displacement, spin and CM memories that we believe were not computed previously.
For the second moment of the news and the corresponding strain, the oscillatory memory terms prove to be larger than the non-oscillatory terms. 
Since the gravitational-wave signal (the strain as a function of time) is the observable that could be measured by gravitational-wave detectors, the oscillatory memory could be the easier to detect.
However, because we compute the PN expressions during the inspiral only (we would need to use numerical relativity waveforms to cover the merger and ringdown phases; see~\cite{Grant:2023jhd}), we do not give quantitative estimates of the detection prospects here, as they would be underestimates due to the truncation of the signal.

The remainder of this paper is organized as follows: 
We review the Bondi-Sachs formalism for vacuum, asymptotically flat spacetimes in Sec.~\ref{sec:BMS formalism}.
In particular, we write the evolution equations for the Bondi metric functions in terms of a set of convenient scalar, charge-like quantities, from which we compute expressions for moments of the news hierarchically. 
The details of these definitions and calculations of the moments of the news are given in Sec.~\ref{sec:Moments of the news}. 
The expressions for the flux contribution to the moments, when expanded in terms of the multipole moments of the shear, are given in Sec.~\ref{sec:MultipoleFlux}. 
In Sec.~\ref{sec:Decompostion of PN expansion}, we compute the first few leading PN-order contributions from the moments of the news and their corresponding contributions to the shear. 
We also show how the PN expansion in~\cite{Blanchet:2013haa} can be understood in terms of contributions from the higher memory effects (moments of the news) in that part.
Next, in Sec.~\ref{sec: PN order of Moments}, we determine the PN orders and contributions to the waveform of the flux and pseudo-flux terms for non-spinning compact binaries in quasi-circular orbits.
We discuss our results and conclude in Sec.~\ref{sec:conclusions}.
A few supplementary results are summarized in four Appendices.

The notation and conventions in this paper are as follows: We use the mostly plus metric signature, and except when explicitly describing the post-Newtonian order of certain terms, we use geometric units with $G=c=1$.
For spacetime coordinate indices, we use Greek letters ($\alpha$, $\beta$, etc.), for spatial indices we use Latin letters (typically) from the middle of the alphabet ($i$, $j$, etc.), and for indices on the two-sphere at infinity we use capital Latin letters ($A$, $B$, etc.).
We also introduce on some quantities a label $\mathrm{I} = \mathrm B$ or $\mathrm E$ to denote the parity of a multipolar quantity (which we put in a font without italics, to avoid confusion with two-sphere indices).
Following a convention from the post-Newtonian literature, we do not always raise or lower spatial indices in the Einstein summation convention.
This convention does not lead to significant ambiguities, as we use a Cartesian coordinate system in which the spatial metric is $\delta_{ij}$ in the context of these PN calculations.

\section{BMS formalism}\label{sec:BMS formalism}

The BMS formalism (see, e.g., the review~\cite{Madler:2016xju}) is frequently used to study asymptotically flat spacetimes, because the coordinates are well adapted to outgoing radiation.
We denote our Bondi-Sachs coordinates by $x^\alpha = \{u,r,\theta^A\}$.
The variable $u$ is the retarded time, and the $u=\mathrm{constant}$ surfaces are null (i.e., the dual vector $k_\alpha = -(\ud u)_\alpha$ is null and normal to these hypersurfaces). 
The angular coordinates are represented here by $\theta^A$, with $A$, $B$ being coordinate indices on the two-sphere.
These angular coordinates are also constant along the null rays to which $k^\alpha$ is tangent: $k^\alpha \partial_\alpha \theta^A=0$. 
Finally, $r$ is an areal radius, in the sense that the area of surfaces of constant $u$ and $r$ is proportional to $r^2$ and independent of $u$. 
The above constraints define Bondi gauge, and they imply
\begin{equation}
    g_{rr}=0, \qquad g_{rA}=0, \qquad \det[g_{AB}]=r^4 q(\theta^A), \label{eqn:Bondi gauge}
\end{equation}
where $q$ is a function of only the coordinates $\theta^A$ on the 2-sphere. 

\subsection{Metric, Einstein equations, boundary conditions, and expansion of the metric}

In this work, we use the following form of the Bondi metric, which is based primarily on the notation of~\cite{Grant:2021hga}:
\begin{align} 
    \ud s^2 =& -\left(1 - \frac{2V}{r}\right) e^{2 \beta/r} \ud u^2 - 2 e^{2 \beta/r} \ud u \ud r 
    \nonumber\\
    &\hspace{1em}+ r^2 \mathcal{H}_{AB} \left(\ud \theta^A - \frac{\mathcal{U}^A}{r^2} \ud u\right) \left(\ud\theta^B - \frac{\mathcal{U}^B}{r^2} \ud u\right).\label{eqn:BMS metric}
\end{align}
The determinant condition in Eq.~\eqref{eqn:Bondi gauge} for this metric can be written in the form $\partial_r \det[\mathcal{H}_{AB}]=0$.

The metric functions are obtained by solving Einstein's equations, subject to initial data on a null hypersurface (and additional boundary data on a prescribed worldtube). 
Einstein's equations in these coordinates take the form of \emph{hypersurface} equations (those that can be solved on surfaces of constant $u$) and \emph{evolution} equations (those that prescribe how to evolve data between different hypersurfaces of constant $u$); see~\cite{Madler:2016xju} for a review.
In vacuum, the hypersurface equations are those arising from the vanishing of the following components of the Ricci tensor: $R^{u}{}_{\alpha}=0$.
These partial differential equations can be solved hierarchically: $\beta$ is determined through a radial integral of terms involving the two-metric $\mathcal H_{AB}$; $\mathcal{U}^A$ is computed from a similar integral involving $\beta$ and $\mathcal H_{AB}$; and $V$ is obtained from an integral that involves the fields $\beta$, $U^A$ and $\mathcal H_{AB}$.
The evolution equations (which come from the components of $R_{AB}=0$ that are trace-free with respect to $\mathcal H_{AB}$) are typically treated as an evolution equation involving the trace-free part of $\partial_u \mathcal H_{AB}$, but they can also be interpreted as a hypersurface equation for this quantity~\cite{Grant:2021hga}. 

The contracted Bianchi identity (specifically the $\beta=u$ and $\beta=A$ components of $\nabla_\alpha {R^\alpha}_\beta=0$ in vacuum) gives rise to the constraints $\partial_r(r^2 e^{\frac{2\beta}{r}}R^r{}_u)=0$ and $\partial_r(r^2 e^{\frac{2\beta}{r}} R^r{}_A)=0$, which imply that the $R^r{}_u$ and $R^r{}_A$ components of the vacuum Einstein equations are satisfied everywhere, if they are satisfied on any finite worldtube or in the limit $r\rightarrow\infty$.
Thus $R^r{}_u = 0$ and $R^r{}_A = 0$ are known as the \emph{supplementary} equations, as compared with the hypersurface and evolution equations, which together are referred to as the \emph{main} equations. 
When the main equations are satisfied, the $\beta=r$ component of the Bianchi identity implies $(\mathcal{H}^{-1})^{AB} R_{AB}=0$, which shows that the trace part of $R_{AB}=0$ is not an independent equation. 
The main and supplementary equations then determine all of the metric functions in the Bondi gauge.

To obtain the metric of an asymptotically flat spacetime, additional boundary conditions on the general metric functions in Eq.~\eqref{eqn:BMS metric} need to be imposed as $r \to \infty$. 
The conditions considered by Bondi and Sachs are given by
\begin{equation} \label{eqn: Asym cond}
    \lim_{r \to \infty} \frac{\beta}{r}=\lim_{r \to \infty} \frac{\mathcal{U}^A}{r}=\lim_{r \to \infty} \frac{V}{r}=0, \qquad \lim_{r \to \infty} \mathcal{H}_{AB}= h_{AB} \, , 
\end{equation}
where $h_{AB}$ is the metric on the unit 2-sphere.
In the expressions throughout the rest of this paper, we will use $h_{AB}$ and $h^{AB}$ to raise and lower indices of tensors with 2-sphere indices, respectively. 
With these boundary conditions, the metric can be expanded in $1/r$, and the hypersurface equations can be solved sequentially to determine $\beta$, $\mathcal{U}^A$ and $V$ in terms of the expansion of $\mathcal{H}_{AB}$ and its derivatives, as well as some functions of integration that do not depend on $r$.
It is then convenient to write $\beta$, $\mathcal{U}^A$ and $V$ in a manner that highlights the leading order behavior in $1/r$ of these functions when subject to the boundary conditions~\eqref{eqn: Asym cond} and the restrictions on $\mathcal{H}_{AB}$ imposed by the determinant condition of Bondi gauge:
\begin{subequations}
\label{eqn: mexp}
\begin{align}
    \beta &= \frac 1r \tilde{\beta}(u,r,\theta^A) , \label{eqn:mexp beta} \\
    V &= m(u,\theta^A) + \frac 1r \mathcal{M}(u,r,\theta^A) , \label{eqn: mexp V} \\
    \mathcal{U}^A &= U^A(u,\theta^B) + \frac 1r V^A(u,\theta^B) + \frac 1{r^2} \Upsilon^A(u,r,\theta^B), \label{eqn: mexp U} \\
    \mathcal{H}_{AB} &= \sqrt{1+\frac{\mathcal{C}_{CD}\mathcal{C}^{CD}}{2r^2}} h_{AB} + \frac{1}{r} \mathcal{C}_{AB}(u,r,\theta^C). \label{eqn: mexp H}
\end{align}
\end{subequations}
Here, $m$ (the \emph{mass aspect}) and $V^A$ are functions of integration of the hypersurface equations for $V$ and $\mathcal{U}^A$, respectively, and thus are functions of $u$ and $\theta^A$. 
The functions $\tilde{\beta}$, $\mathcal{M}$, \emph{and} $\Upsilon^A$ depend on all four coordinates, and they admit an expansion in $1/r$ starting at $O(1)$; the coefficients in this expansion can be written in terms of the expansion of $\mathcal{H}_{AB}$, its derivatives, and the other functions of integration.

The determinant condition [which determines the form of Eq.~\eqref{eqn: mexp H}] implies that $\mathcal H_{AB}$, the angular part of the metric, can be written entirely in terms of a symmetric, trace-free\footnote{Here, and below, when we write ``trace-free,'' we mean with respect to $h_{AB}$.} (STF) tensor $\mathcal{C}_{AB}$. 
The $1/r$ part of the expansion of $\mathcal{C}_{AB}$ vanishes in vacuum~\cite{Flanagan:2015pxa}, and we write the expansion of it in powers of $1/r$ using the notation of~\cite{Grant:2021hga}: 
\begin{equation}
    \mathcal{C}_{AB} = C_{AB}(u,\theta^C) + \frac{1}{r^2}\sum_{n = 0}^\infty \frac{1}{r^n} \ord{n}{\mathcal E}_{AB}(u,\theta^C) \, . \label{eqn: mexp shear}
\end{equation}
The coefficients $\ord{n}{\mathcal E}_{AB}$ in the expansion (which, following~\cite{Compere:2022zdz}, we refer to as the ``higher Bondi aspects'') evolve with retarded time according to evolution equations; however, the evolution of the leading order term $C_{AB}$ (the \emph{shear}) remains unconstrained. 
The time derivative of the shear can be freely specified, and it is known as the Bondi \emph{news tensor} $N_{AB}=\partial_u C_{AB}$. 

In the literature, the integration constant $V^A$ is conventionally written in terms of $N^A$ (the \emph{angular momentum aspect}), defined as
\begin{equation}
    N^A(u,\theta^B) \equiv -\frac{3}{2}V^A + \frac{3}{32} \mathscr{D}^A(C_{BC}C^{BC}) + \frac{3}{4} C^{AB} \mathscr{D}^C C_{BC}, \label{eqn:N}
\end{equation}
where $\mathscr{D}_A$ is the covariant derivative on the two-sphere, which is compatible with the metric $h_{AB}$. 
Lastly, the time evolution of the functions of integration, the mass aspect and the angular momentum aspect, are computed from the leading-order terms in $1/r$ of the supplementary equations. 
One can then asymptotically solve the hypersurface and the evolution equations order by order in $1/r$ and the entire BMS metric is then determined by the mass aspect $m(u,\theta^A)$, the angular momentum aspect $N^A(u,\theta^A)$, the shear $C_{AB}(u,\theta^A)$ and the higher Bondi aspects $\ord{n}{\mathcal E}_{AB} (u,\theta^A)$.

\subsection{Evolution equations for the expanded Bondi metric functions}

The news tensor plays an important role in determining the evolution of the Bondi metric functions.
The three lowest-order evolution equations can be written as (see~\cite{Grant:2021hga,Madler:2016xju}):
\begin{subequations}
\label{eqn:Evol}
\begin{align}
    \partial_u m &= \frac{1}{4} \left(\mathscr{D}_A \mathscr{D}_B N^{AB} - \frac{1}{2} N_{AB} N^{AB}\right), \label{eqn:dot_m} \\
    \partial_u N_A &= \mathscr{D}_A m - \frac{1}{4} \epsilon_{BA} \mathscr{D}^B (\epsilon^{CD} \mathscr{D}_C \mathscr{D}^E C_{DE}) + \frac{1}{4} \left(N^{BC} \mathscr{D}_B C_{CA} + 3 C_{AB}\mathscr{D}_C N^{BC}\right), \label{eqn:dot_N} \\
    \partial_u \ord{0}{\mathcal E}_{AB} &= \frac{1}{4} N_{CD} C^{CD} C_{AB} + \frac{1}{3} \STF \mathscr{D}_A N_B + \frac{1}{4} C_B{}^C \mathscr{D}_{[A} \mathscr{D}^D C_{C]D} + \frac{1}{2} m C_{AB}. \label{eqn:dot_E0}
\end{align}
\end{subequations}
A schematic form of the evolution equations for the higher Bondi aspects with $n > 1$ is given in Eq.~(4.29) of~\cite{Grant:2021hga}.
The $\STF$ operator above means to take the symmetric, trace-free part of the tensor on the free indices using the metric $h_{AB}$. 

In this paper, we will work with a modified angular momentum aspect given by 
\begin{subequations}
\begin{align}
    \hat{N}^A & = N^A - \mathcal W^A , \\
    \mathcal{W}^A & \equiv \frac{1}{8}(C_B{}^C \mathscr{D}_C C^{AB} + 3 C^{AB}\mathscr{D}^C C_{BC}).
\end{align}
\end{subequations}
One reason for this definition is that $\hat N^A$ simplifies the integrand for the ``conserved quantity'' conjugate to the Lorentz symmetries of the BMS algebra at null infinity, as defined by Wald and Zoupas~\cite{Wald:1999wa} (see also~\cite{Flanagan:2015pxa}).\footnote{
Other definitions of these conserved quantities have been discussed in the literature~\cite{Compere:2019gft}, and there are various arguments for and against them~\cite{Elhashash:2021iev, Grant:2021sxk, Chen:2022fbu}.}
This definition will also be useful for comparing our PN calculations that we will perform in Sec.~\ref{sec: PN order of Moments} with prior PN calculations of the spin and CM memory effects.

Many of the calculations in this paper involving the evolution equations~\eqref{eqn:Evol} will be simplified by introducing scalar versions of the metric functions $\hat N^A$ and $\ord{n}{\mathcal E}_{AB}$.
We first define
\begin{equation}
    Q_0 \equiv m, \qquad Q_1\equiv \mathscr{D}_A \hat N^A, \qquad Q_{n+2} \equiv \mathscr{D}_A \mathscr{D}_B \ord{n}{\mathcal E}^{AB}. \label{eqn:Def Elec}
\end{equation}
The vector $\hat N^A$ and STF tensors $\ord{n}{\mathcal E}_{AB}$ each contain two degrees of freedom, and taking the divergences extracts the ``electric'' or curl-free parts of these vectors and tensors. 
The corresponding ``magnetic'' or divergence-free parts of the tensors can be obtained from 
\begin{equation}
  Q_0^*\equiv - \frac{1}{4}\epsilon^{CA}\mathscr{D}_A \mathscr{D}^B C_{CB},\qquad  Q_1^* \equiv \epsilon_{BA} \mathscr{D}^A \hat N^B, \qquad Q_{n+2}^* \equiv \epsilon_{BA} \mathscr{D}_C \mathscr{D}^A \ord{n}{\mathcal E}^{BC}. \label{eqn:Def Mag}
\end{equation}
That $Q_1$ and $Q_1^*$ capture all of the degrees of freedom in $\hat N^A$, say, follows from the fact that vectors on the sphere can be written in terms of a curl-free and divergence-free part; a similar result holds for $Q_{n + 2}$, $Q_{n + 2}^*$, and $\ord{n}{\mathcal E}_{AB}$.
Equivalently, Eq.~\eqref{eqn:Def Mag} can be written as
\begin{equation}
   Q_0^*\equiv -\frac{1}{4}\mathscr{D}_A \mathscr{D}_B (^*C)^{AB}, \qquad Q_1^* \equiv \mathscr{D}_A (^* \hat N)^A, \qquad Q_{n+2}^* \equiv \mathscr{D}_A \mathscr{D}_B (^* \ord{n}{\mathcal E})^{AB},
\end{equation}
where we have defined the (left) dual of a tensor field on the sphere by
\begin{equation}
    (^* T)_{A_1 \cdots A_s} \equiv \epsilon^B{}_{A_1} T_{BA_2 \cdots A_s}.
\end{equation}
It can be shown (for example, using Proposition~D.1 of~\cite{Grant:2021sxk}) that if a tensor field $T_{AB}$ is STF, its dual is also STF.

The evolution equations for these scalar metric functions (or ``charges'') take the following form:
\begin{subequations} \label{eqn:evol_Q}
\begin{align} 
  \dot Q_0 &= \frac{1}{4} \mathscr D_A \mathscr D_B N^{AB} + \mathcal F_0, \label{eqn: evol Q_0}\\
  \dot Q_0^* &= - \frac{1}{4} \mathscr D_A \mathscr D_B (^*N)^{AB} +\mathcal F^*_0, \label{eqn: evol Q_0 mag}\\
  \dot Q_1 &=  \mathscr{D}^2 Q_0  + \mathcal{F}_{1} , \label{eqn: evol Q_1 elec}\\
  \dot Q^*_1 &= \mathscr{D}^2 Q_0 ^* + \mathcal F^{*}_1 , \label{eqn: evol Q_1 mag}\\
  \dot Q_2&= \frac{1}{6}(\mathscr D^2+2) Q_1 +\mathcal F_2 + \mathcal G_2^{\mathrm{rad}} + \mathcal G_2^{\mathrm{nonrad}},\label{eqn: evol Q_2 elec}\\
  \dot Q_2^*&= \frac{1}{6}(\mathscr D^2+2) Q_1 ^* +\mathcal F^*_2 + \mathcal G_2^{^*\mathrm{rad}} + \mathcal G_2^{^*\mathrm{nonrad}}.\label{eqn: evol Q_2 mag}
\end{align}
\end{subequations}
In the expressions above, we have introduced a notation for the \emph{fluxes} $\mathcal F_n$ as well as \emph{radiative and non-radiative ``pseudo-fluxes''} $\mathcal G_n^{\mathrm{rad}}$ and $\mathcal G_n^{\mathrm{nonrad}}$, respectively.
The expression for the fluxes $\mathcal F_n\leq 2$ are given by
\begin{subequations}
\begin{align}
    \mathcal{F}_0&= -\frac{1}{8}N_{AB}N^{AB},\qquad  \qquad \mathcal F^*_0=0,\label{eqn: F0}  \\
    \mathcal{F}_1&= \frac{1}{8} \mathscr{D}^A \left[(N^{BC}\mathscr{D}_B C_{CA}-C^{BC}\mathscr{D}_B N_{CA}) + 3 (C_{AB}\mathscr{D}_C N^{BC}- N_{AB}\mathscr{D}_C C^{BC})\right], \label{eqn:F1 elec} \\
    \mathcal F_1^* &=\frac{1}{8} \mathscr{D}^A \left[N^{BC}\mathscr{D}_B (^* C)_{CA} - C^{BC}\mathscr{D}_B (^*N)_{CA}+ 3 (^*C_{AB}\mathscr{D}_C N^{BC} - {}^*N_{AB} \mathscr{D}_C C^{BC})\right], \label{eqn:F1 mag} \\
\mathcal{F}_2&=\frac{1}{4} \mathscr{D}_A \mathscr{D}_B (N_{CD}C^{CD} C^{AB}), \label{eqn: F2 elec} \\
 \mathcal{F}_2 ^* &= \frac{1}{4} \mathscr{D}_A \mathscr{D}_B [N_{CD}C^{CD} (^* C)^{AB}]. \label{eqn:F2 mag} 
\end{align}
\end{subequations}
The pseudo-fluxes for $n=2$ are given by
\begin{subequations}
\begin{align}
    &\mathcal{G}_2 ^{\mathrm{rad}}= \frac{1}{8} \mathscr{D}_A \mathscr{D}_B \{[\mathscr{D}_{E} \mathscr{D}_{F} (^* C)^{EF}] (^* C)_{AB}\} +  \frac{1}{6}(\mathscr{D}^2+2) \mathscr{D}_A\mathcal{W}^A, \label{eqn: G rad2 elec}\\
    &\mathcal{G}_2 ^{\mathrm{nonrad}}= \frac{1}{2}\mathscr{D}_A \mathscr{D}_B (m C^{AB}),\label{eqn: G nonrad2 elec}\\
    &\mathcal{G}_2 ^{\mathrm{rad}\;*} = \frac{1}{8} \mathscr{D}_A \mathscr{D}_B \{[\mathscr{D}_{E} \mathscr{D}_{F} (^* C)^{EF}] C_{AB}\} + \frac{1}{6}(\mathscr{D}^2+2) \mathscr{D}_A(^* \mathcal{W}^A),\label{eqn: G rad2 mag}\\
     &\mathcal{G}_2 ^{\mathrm{nonrad}\;^*}= \frac{1}{2}\mathscr{D}_A \mathscr{D}_B [m (^* C)^{AB}].\label{eqn: G nonrad2 mag}
\end{align}
\end{subequations}

In general for all $n > 0$, the evolution-type Einstein equations can be written in a scalar, flux-balance form as\footnote{While this holds explicitly for $n \leq 2$ by Eq.~\eqref{eqn:evol_Q}, it is not immediately obvious that the two differential operators which appear in Eqs.~\eqref{eqn: evol Q_n elec} and~\eqref{eqn: evol Q_n mag} are necessarily the same for all $n$.  However, this can be proven using the schematic form of the evolution equations in Eq.~(4.29) of~\cite{Grant:2021hga}, specifically the fact that the differential operator that appears in that equation commutes with the action of taking the dual.}
\begin{subequations} \label{eq:Q_n_PDEs}
\begin{align} 
  \dot Q_n &= \mathcal D_n Q_{n - 1} + \mathcal F_n + \mathcal G_n, \label{eqn: evol Q_n elec}\\
  \dot Q^*_{n} &= \mathcal D_{n} Q^*_{n - 1} + \mathcal F^*_{n } + \mathcal G^*_{n} .  \label{eqn: evol Q_n mag}
\end{align}
\end{subequations}
where $\mathcal G_n$ and $\mathcal G_n^*$ contain both radiative and non-radiative parts generically.
The fluxes $\mathcal{F}_n$ are nonlinear terms that vanish in the absence of radiation. 
The $\mathcal{G}_n$ terms are also non-linear, but they can be nonzero in a non-radiative spacetime, which is why we refer to them as \emph{pseudo-fluxes}. 
The operators $\mathcal{D}_n$ are functions of covariant derivatives on 2-sphere, whose explicit form for $n\leq2$ can be inferred from the expressions above.
In general, Eq.~(2.24) of~\cite{Grant:2021hga} (note the difference in notation!) implies that $\mathcal D_n \propto \mathscr D^2 + n (n - 1)$, so that when acting on a spherical harmonic $Y_{lm}$ it annihilates $l = n - 1$:
\begin{equation} \label{eqn:annihilator}
    \mathcal D_n Y_{(n - 1)m} = 0.
\end{equation}
We will not use the evolution equations with $n\geq 2$, which is why we do not give their explicit forms here.

\section{Moments of the news and higher memory effects} \label{sec:Moments of the news}

In this section, we introduce and compute the moments of the news in terms of changes in charges, as well as integrals of fluxes and pseudo-fluxes.
We will use these moments to characterize the higher memory effects of this paper, though more generally, these moments could be used to characterize the full Bondi news as a function of time, much like the moments of a probability distribution. 
As in~\cite{Grant:2023jhd}, we use somewhat different moments than those introduced in~\cite{Grant:2021hga}, because the moments in~\cite{Grant:2021hga} were those that could be related straightforwardly to the curve deviation of~\cite{Flanagan:2018yzh}.
In this work, we are more interested in relating the moments of the news to a portion of the gravitational-wave strain.
For this purpose, it is useful to define the moments in terms of a multiple integral over the news: 
\begin{equation} \label{eqn:Mom news}
    \begin{split}
        \ord{n}{\mathcal N}_{AB} (u_1, u_0) &= \int^{u_1} _{u_0} \ud u_2 \cdots \int^{u_{n+1}}_{u_0}\ud u_{n+2}\, N_{AB}(u_{n+2}) \\
        &\equiv \left(\prod_{m = 1}^{n + 1} \int_{u_0}^{u_m} \ud u_{m + 1}\right) N_{AB} (u_{n + 2}),
    \end{split}
\end{equation}
where the second line in this equation defines a repeated composition notation for a series of operators (in this case, integration).\footnote{Note that, in this product notation, we assume that successive terms appear left-to-right (in this case, the first term in the product is the outer-most integral). Since composition of operators proceeds right-to-left, this implies that the operators which appear are applied in the reverse order that they appear in the product.}
The moments in Eq.~\eqref{eqn:Mom news} can be related to a (finite) linear combination of the moments defined in~\cite{Grant:2021hga} for each $n$; the relationship between the two is given in~\ref{app:MomentRelations}.

\subsection{Procedure for computing the moments of the news}

In~\cite{Grant:2021hga}, the moments of the news were obtained by defining charges from the Bondi aspects and integrating the evolution-type Einstein equations in Eq.~\eqref{eqn:Evol} (or the equivalents for the higher Bondi aspects). 
We perform a similar calculation for our scalar aspects $Q_n$, which we will summarize below for the different moments.

\subsubsection{Zeroth moment}

The zeroth moment is the simplest and best understood case, though it also is a special case.
Specifically, the zeroth moment $\ord{0}{\mathcal N}_{AB}$ is the change in the shear that occurs after gravitational waves pass, which is precisely the displacement memory. 
It can be computed by integrating Eq.~\eqref{eqn: evol Q_0} once over retarded time:
\begin{equation} \label{eqn:zeroth moment}
     \frac{1}{4}\ord{0}{\mathcal N} (u, u_0) = \Delta {Q}_0 (u, u_0) - \int_{u_0} ^{u} \ud u_1 {\mathcal{F}_0}(u_1),
\end{equation}
where, for all $n\geq0$, we define $\Delta Q_n(u_1, u_0)$ and scalar moments $\ord{n}{\mathcal N}$ by
\begin{align}
    \Delta Q_n(u_1, u_0) &\equiv Q_n(u_1) - Q_n(u_0).\\
    \ord{n}{\mathcal N} (u, u_0)&\equiv\mathscr{D}_A \mathscr{D}_B \ord{n}{\mathcal N}^{AB} (u, u_0).
\end{align}
Equation~\eqref{eqn:zeroth moment} has a charge contribution $\Delta{Q}_0$ which is conserved in the absence of radiation and a flux contribution $\mathcal{F}_0$.
The flux contribution is equivalent to the nonlinear contribution to the memory computed by Christodoulou~\cite{Christodoulou:1991cr}.

Our expression for the zeroth moment contains just the electric-parity piece, because the differential operator acting on the news in Eq.~\eqref{eqn: evol Q_0 mag} removes the magnetic-parity part.
The magnetic-part of the zeroth moment of the news (that is, the difference in the shear) can be obtained from taking the difference of Eq.~\eqref{eqn: evol Q_1 mag} at two different retarded times (one before and one after the waves pass), as the magnetic part of the shear appears on the right-hand side.
The flux term vanishes (because the news vanishes when there are no gravitational waves), so the magnetic part is determined $\Delta \dot Q^*_1$.
Because we focus on the flux terms in this paper, we will not compute it here.\footnote{Note that it has been shown in~\cite{Satishchandran:2019pyc} that there are sources of stress-energy that satisfy standard energy conditions that produce a non-zero magnetic-parity displacement memory (i.e., zeroth moment of the news). These ``vector'' types of memory in~\cite{Satishchandran:2019pyc} would then arise from having spacetimes with a non-vanishing change in the time derivative of the charge term.}

\subsubsection{First moment}

For obtaining the electric part of the first moment, we integrate Eq.~\eqref{eqn: evol Q_0} twice to obtain
\begin{equation} \label{eqn:int 0 moment}
    \frac{1}{4} \ord{1}{\mathcal N} (u, u_0) = \int_{u_0} ^{u} \ud u_1 \Delta {Q}_0 (u_1,u_0) - \int_{u_0} ^{u} \ud u_1 \int_{u_0} ^{u_1} \ud u_2 \mathcal{F}_0(u_2),
\end{equation}
and similarly for the magnetic part we need to integrate Eq.~\eqref{eqn: evol Q_0 mag} twice:
\begin{equation}
    \frac{1}{4}\ord{1}{\mathcal N^*} (u, u_0) = \int_{u_0} ^{u} \ud u_1 \Delta {Q}_0^*(u_1,u_0), \label{eqn:int 0 moment mag}
\end{equation}
where we have defined a magnetic scalar moment by
\begin{align}
    \ord{n}{\mathcal N^*} (u, u_0)&\equiv -\mathscr{D}_A \mathscr{D}_B (^*\ord{n}{\mathcal N})^{AB} (u, u_0).
\end{align}
Thus, we will need expressions for the integrals of $\Delta Q_0$ and $\Delta Q^* _0$ to obtain a similar expression for first moment in terms of changes of charges and fluxes (as we have for the zeroth moment). 
We can derive such an expression by defining new scalar charges as follows:
\begin{subequations}
\begin{align}
    \tilde{{Q}}_1(u,\tilde{u}) &\equiv {Q}_1(u)+ (\tilde{u}-u)\mathscr{D}^2 {Q}_0(u) , \\
    \tilde{{Q}}^*_1(u,\tilde{u}) &\equiv {Q}^*_1(u)+ (\tilde{u}-u)\mathscr{D}^2 {Q}^*_0(u) .
\end{align}
\end{subequations}
Then the difference of the charges is given by
\begin{align}
    \Delta\tilde{Q}_1(u,u_0;\tilde{u}) &\equiv \tilde{Q}_1(u,\tilde{u})-\tilde{Q}_1(u_0,\tilde{u})\nonumber\\
    &={Q}_1(u)-Q_1(u_0) +(\tilde{u}-u)\mathscr{D}^2  Q_0(u)- (\tilde{u}-u_0)\mathscr{D}^2  Q_0(u_0) , 
\end{align}
and if we consider the special case of $\tilde{u}=u$, the expression reduces to 
\begin{align} \label{eq:DeltaQtilde1}
    \Delta \tilde Q_1 (u, u_0) \equiv \Delta\tilde{Q}_1(u,u_0;u) = Q_1(u)-Q_1(u_0)- (u-u_0)\mathscr{D}^2  Q_0(u_0) .
\end{align}
The evolution of $\Delta\tilde{Q}_1(u,u_0)$ can be computed by differentiating Eq.~\eqref{eq:DeltaQtilde1} with respect to $u$ and using Eq.~\eqref{eqn: evol Q_1 elec} to eliminate $\dot Q_1$ from the expression: 
\begin{align} \label{eqn: evol of change in tQ1}
    \partial_u \Delta \tilde{Q}_1(u,u_0)=\mathcal{F}_1 + \mathscr{D}^2 \Delta Q_0(u,u_0) .
\end{align}
Note that on the right-hand side of Eq.~\eqref{eqn: evol of change in tQ1}, there is the term proportional to $\mathscr{D}^2 \Delta Q_0(u,u_0)$.
If we apply the operator $\mathscr D^2$ to Eq.~\eqref{eqn:int 0 moment}, Eq.~\eqref{eqn: evol of change in tQ1} can then be used to replace the integrand involving $\Delta Q_0$ with a flux and a time derivative of a charge. 
Integrating Eq.~\eqref{eqn: evol of change in tQ1} and substituting it into $\mathscr D^2$ applied to Eq.~\eqref{eqn:int 0 moment} gives
\begin{align} \label{eqn:first moment elec}
    \frac{1}{4}\mathscr{D}^2 \ord{1}{\mathcal N}(u, u_0) = \Delta \tilde{Q}_1(u, u_0) -\int_{u_0} ^{u} \ud u_1 \mathcal{F}_1(u_1) - \mathscr{D}^2 \int_{u_0} ^{u} \ud u_1 \int_{u_0} ^{u_1} \ud u_2 \mathcal{F}_0(u_2) .
\end{align}
These steps work similarly for the magnetic part of the moment of the news, which can be written as 
\begin{align} \label{eqn:first moment mag}
    \frac{1}{4} \mathscr{D}^2 \ord{1}{\mathcal N^*}(u, u_0) = -\Delta \tilde{Q}_1 ^*(u, u_0) +\int_{u_0} ^u \ud u_1\tilde{\mathcal F}_1 ^* (u_1). 
\end{align}
This calculation for the electric (respectively, magnetic) part of the first moment of the news has used the evolution equation for $Q_1$ ($Q_1^*$) and a redefinition of the charge to recast the zeroth charge $\Delta\tilde{Q}_0$ $(\Delta\tilde{Q}_0^*)$ as a total derivative piece $\partial_u\Delta \tilde{Q}_1$ $(\partial_u\Delta \tilde{Q}_1^*)$ and a flux piece $\mathcal{F}_1$ $(\mathcal{F}^*_1)$. 
The electric part of the first moment of the news now has three parts, a total change in the charge, an integral over a ``new'' flux ($\mathcal{F}_1$) and a double integral over the flux piece of the zeroth moment; it is closely related to the CM memory effect in~\cite{Nichols:2018qac}. 
The magnetic part of the first moment is two integrals of the dual of the news, and the left-hand side of Eq.~\eqref{eqn:first moment mag} is equivalent to the spin memory effect (see~\cite{Nichols:2017rqr}).

The procedure used to compute a single integral of $\Delta\tilde{Q}_0$ can be generalized to compute $n$ integrals, as we describe in the next part.

\subsubsection{General procedure for the \textnormal{n}th moment}

To compute the $n$th moment of the news in terms of changes in charges and integrals of fluxes, we first define the modified electric and magnetic-parity scalars:
\begin{equation} \label{eqn:tilde_Q}
  \tilde Q_n (u, \tilde u) \equiv \sum_{m = 0}^n \left(\prod_{k = 0}^{n - m - 1} \frac{\tilde u - u}{k + 1} \mathcal D_{n - k}\right) Q_m (u),
\end{equation}
where, like in Eq.~\eqref{eqn:Mom news}, we are using a product notation for repeated composition of linear operators.
We apply the convention that when a product is empty (e.g., when the lower bound is higher than the upper bound), the product becomes the identity, and so the first term in this sum is simply $Q_n (u)$.
Similarly, one can define
\begin{equation} \label{eqn:tilde_Q mag}
  \tilde Q_n ^*(u, \tilde u) \equiv \sum_{m = 0}^n \left(\prod_{k = 0}^{n - m - 1} \frac{\tilde u - u}{k + 1} \mathcal D_{n - k}\right) Q_m^* (u).
\end{equation}

We will focus on the electric-parity scalars $\tilde Q_n$, as the discussion for the magnetic-parity scalars will be completely analogous. 
The quantity that will be relevant for computing moments of the news will be differences in these modified scalars; thus, it is convenient to define
\begin{equation}
  \Delta \tilde Q_n (u_1, u_0; \tilde u) \equiv \tilde Q_n (u_1, \tilde u) - \tilde Q_n (u_0, \tilde u).
\end{equation}
For computing the moments of the news, we will need the special case when $\tilde u = u_1$, in which case $\Delta \tilde Q_n (u_1, u_0) \equiv \Delta \tilde Q_n (u_1, u_0; u_1)$ can be written as
\begin{equation} \label{eqn:Delta_tilde_n}
  \begin{split}
    \Delta \tilde Q_n (u_1, u_0) &= Q_n (u_1) - \tilde Q_n (u_0, u_1)  \\
    &= Q_n (u_1) - Q_n (u_0) - \sum_{m = 0}^{n - 1} \left(\prod_{k = 0}^{n - m - 1} \frac{u_1 - u_0}{k + 1} \mathcal D_{n - k}\right) Q_m (u_0).
  \end{split}
\end{equation}
The first two terms in the second line on the right-hand side of Eq.~\eqref{eqn:Delta_tilde_n} are simply the integral of the derivative of $Q_n(u)$ from $u_0$ to $u_1$.
The last term can be written as an integral from $u_0$ to $u_1$ by integrating each term in the sum in Eq.~\eqref{eqn:tilde_Q} with respect to the second argument:
\begin{equation}
    \int_{u_0}^{u_1} \tilde Q_{n - 1} (u_0, u) \ud u = \sum_{m = 0}^{n - 1} \frac{u_1 - u_0}{n - m} \left(\prod_{k = 0}^{n - m - 2} \frac{u_1 - u_0}{k + 1} \mathcal D_{n - 1 - k}\right) Q_m (u_0).
\end{equation}
By applying $\mathcal D_n$ to this equation, one can absorb the term outside of the product by changing the upper bound of the product to $n - m - 1$, obtaining the third term on the right-hand side of Eq.~\eqref{eqn:Delta_tilde_n}.
This means that we can write $\Delta \tilde Q_n (u_1, u_0)$ as the following integral:
\begin{equation}
  \Delta \tilde Q_n (u_1, u_0) = \int_{u_0}^{u_1} \left[\dot Q_n (u) - \mathcal D_n \tilde Q_{n - 1} (u_0, u) \right] \ud u.
\end{equation}
Integrating Eq.~\eqref{eqn: evol Q_n elec} and using the first line of Eq.~\eqref{eqn:Delta_tilde_n}, we therefore find that
\begin{equation} \label{eq:DeltaQnTildeRecur}
  \Delta \tilde Q_n (u_1, u_0) = \int_{u_0}^{u_1} \left[\mathcal D_n \Delta \tilde Q_{n - 1} (u, u_0)  + \mathcal F_n (u) + \mathcal G_n (u)\right] \ud u.
\end{equation}
Equation~\eqref{eq:DeltaQnTildeRecur} can then be applied recursively until the expression for $\Delta \tilde Q_n (u_1, u_0)$ can be written as a linear combination of integrals of fluxes and pseudo-fluxes and $n$ integrals of $\Delta Q_0 (u_1, u_0)$:
\begin{equation} \label{eq:DeltaQnQ0}
  \begin{split}
    \Delta \tilde Q_n (u_1, u_0) &= \left(\prod_{m = 0}^{n - 1} \mathcal D_{n - m} \int_{u_0}^{u_{m + 1}} \ud u_{m + 2}\right) \Delta Q_0 (u_{n + 1}, u_0) \\
    &\hspace{1em}+ \sum_{m = 0}^{n - 1} \int_{u_0}^{u_1} \ud u_2 \left(\prod_{k = 0}^{m - 1} \mathcal D_{n - k} \int_{u_0}^{u_{k + 2}} \ud u_{k + 3}\right) \left[\mathcal F_{n - m} (u_{m + 2}) + \mathcal G_{n - m} (u_{m + 2})\right],
  \end{split}
\end{equation}
To write this in terms of the $n$th moment of the news, we use Eq.~\eqref{eqn:zeroth moment}, which when substituted into Eq.~\eqref{eq:DeltaQnQ0} yields
\begin{equation} \label{eqn:moment_of_news elec}
  \begin{split}
    \frac{1}{4} \left(\prod_{m = 0}^{n - 1} \mathcal D_{n - m}\right) \ord{n}{\mathcal N} (u_1, u_0) = \Delta \tilde Q_n (u_1, u_0) - \sum_{m = 0}^n &\int_{u_0}^{u_1} \ud u_2 \left(\prod_{k = 0}^{m - 1} \mathcal D_{n - k} \int_{u_0}^{u_{k + 2}} \ud u_{k + 3}\right) \\
    &\times \left[\mathcal F_{n - m} (u_{m + 2}) + \mathcal G_{n - m} (u_{m + 2})\right],
  \end{split}
\end{equation}
where, for brevity, we have defined $\mathcal G_0\equiv \mathcal G_1 \equiv 0$.
Expressions involving the magnetic scalars as the dual of moments of the news can be calculated from integrals of $\Delta Q_0 ^*$ in a completely analogous way, and have exactly the same form, apart from a minus sign:
\begin{equation} \label{eqn:moment_of_news mag}
  \begin{split}
    \frac{1}{4} \left(\prod_{m = 0}^{n - 1} \mathcal D_{n - m}\right) \ord{n}{\mathcal N}^* (u_1, u_0) = -\Delta \tilde Q_n^* (u_1, u_0) + \sum_{m = 0}^n &\int_{u_0}^{u_1} \ud u_2 \left(\prod_{k = 0}^{m - 1} \mathcal D_{n - k} \int_{u_0}^{u_{k + 2}} \ud u_{k + 3}\right) \\
    &\times \left[\mathcal F_{n - m}^* (u_{m + 2}) + \mathcal G_{n - m}^* (u_{m + 2})\right] .
  \end{split}
\end{equation}
We similarly defined $\mathcal G_0^*\equiv \mathcal G_1^* \equiv 0$.
Since $\mathcal F_0^* = 0$ as well, for calculations the upper limit of the sum can be taken as $n - 1$.
Equations~\eqref{eqn:moment_of_news elec} and~\eqref{eqn:moment_of_news mag} then can be applied to any moment of the news once the corresponding fluxes, pseudo-fluxes, and charges are determined. 

\subsubsection{Procedure applied to the second moment}

As an example, we give the expressions for the second moment and its dual, as given by Eqs.~\eqref{eqn:moment_of_news elec} and~\eqref{eqn:moment_of_news mag}:
\begin{subequations}{\label{eqn:Second_moment}}
\begin{align}
    \frac{1}{24} (\mathscr{D}^2 +2)\mathscr{D}^2 \ord{2}{\mathcal N} (u, u_0) &= \Delta \tilde{Q}_2(u, u_0) - \int_{u_{0}} ^u \ud u_1 \bigg[\mathcal{F}_2(u_1) + \mathcal{G}_2 ^{\mathrm{rad}}(u_1) + \mathcal{G}_2 ^{\mathrm{nonrad}}(u_1)\bigg] \nonumber\\
    &\hspace{1em}-\frac{1}{6} (\mathscr{D}^2+2) \int_{u_{0}} ^u \ud u_1\int_{u_{0}} ^{u_1}   \ud u_2 \left[\mathcal{F}_1(u_2) + \mathscr{D}^2 \int_{u_{0}} ^{u_2} \ud u_3 \mathcal{F}_0(u_3)\right], \label{eqn: second moment elec}
\end{align}
\begin{align}
        \frac{1}{24} (\mathscr{D}^2 +2)\mathscr{D}^2 \ord{2}{\mathcal N^*}(u, u_0) &= -\Delta \tilde{Q}_2 ^*(u, u_0)+ \int_{u_{0}} ^u \ud u_1 [\mathcal{F}_2 ^* (u_1) + \mathcal{G}_2 ^{\mathrm{rad}\;*} (u_1)
        +\mathcal{G}_2 ^{\mathrm{nonrad}\;*} (u_1) \nonumber ] \nonumber \\
        &\hspace{1em}+\frac{1}{6} (\mathscr{D}^2+2) \int_{u_{0}} ^u \ud u_1\int_{u_{0}} ^{u_1}   \ud u_2 \mathcal{F}_1 ^* (u_2).  \label{eqn:second moment mag}
\end{align}
\end{subequations}
Unlike the zeroth or first moments, the second moment of the news and its dual have radiative and non-radiative pseudo-flux terms in addition to the charge and the flux pieces. 

The three memory effects (displacement, spin, and CM) appear in the zeroth and the first moments of the news.
The second and higher moments of the news contain a hierarchy of memory effects that have only been calculated for astrophysical sources very recently~\cite{Grant:2023jhd}.
In this work, we calculate general multipolar expressions and PN results for the second moment of the news (i.e., the ballistic memory).
We could have extended the calculations to higher moments, though we do not do so in this paper.

\subsection{Computing the moments of the news and the corresponding shear} \label{subsec:shear-from-moments}

While the equations~\eqref{eqn:moment_of_news elec} and~\eqref{eqn:moment_of_news mag} have related the moments of the news to changes in charges and fluxes, the fluxes depend on the news, which make these equations integral equations for the moments. 
These expressions are then equivalent, in a sense, to the partial differential equations in Eq.~\eqref{eq:Q_n_PDEs}, which were used to derive the expressions~\eqref{eqn:moment_of_news elec} and~\eqref{eqn:moment_of_news mag}.
Thus, without making further assumptions, the integral expressions for the moments do not ``determine'' the moments of the news from the charges and fluxes; they are consistency relationships for solutions of Einstein's equations.

However, there are often scenarios (most frequently for analytical approximation methods) in which the equations for the moments of the news can be used to calculate a contribution to a moment from the changes in charges or fluxes that involve approximate gravitational waveforms that do not contain this moment (at a given order in a particular approximation method, for example).
Some of these contexts include in post-Newtonian theory (see, e.g.,~\cite{Favata:2008yd,Blanchet:1992br,Nichols:2017rqr,Compere:2019gft}), linear perturbations of black holes~\cite{Burko:2020gse}, and even some numerical relativity waveforms computed using extrapolation methods at finite radii (see, e.g.,~\cite{Favata:2009ii,Nichols:2017rqr,Talbot:2018sgr,Mitman:2020bjf,Mitman:2020pbt,Grant:2023jhd}).\footnote{Numerical relativity calculations using Cauchy-characteristic evolution or matching do not have this property.
In these cases, the moments of the news can be used as a check of the numerical accuracy of the solutions.}

We will use the expressions for the moments of the news in this latter sense, where we will treat the charges and fluxes as known quantities and compute a moment of the news that is required to be consistent with the changes in the charges and the integral of the fluxes and pseudo-fluxes.
Our focus in this paper, moreover, will be on the flux and pseudo-flux (rather than the charge) contribution, and the ``new'' part of the flux: namely, that which is not a further integral of a flux (or pseudo-flux) that appears in a lower moment of the news. 

Because gravitational-wave interferometers (such as LIGO, Virgo and KAGRA) are calibrated to measure the strain rather than its time integrals, it is useful to have an observable coming from the moment of the news that is related to the gravitational-wave strain.
Given our definition of the moments of the news, if instead
we take $n$ derivatives of both sides of Eqs.~\eqref{eqn:moment_of_news elec} and~\eqref{eqn:moment_of_news mag}, we obtain
\begin{equation} \label{eqn:shear_contribution elec}
  \begin{split}
    \frac{1}{4} &\left(\prod_{m = 0}^{n - 1} \mathcal D_{n - m}\right) \mathscr D_A \mathscr D_B [C^{AB} (u) - C^{AB} (u_0)] \\
    &= \frac{\partial^n}{\partial u^n} \Delta \tilde Q_n (u, u_0) - \sum_{m = 0}^{n - 1} \left(\prod_{k = 0}^{m - 1} \mathcal D_{n - k}\right) \left(\frac{\ud}{\ud u}\right)^{n - m - 1} \left[{\mathcal F}_{n - m} (u) + {\mathcal G}_{n - m} (u)\right] \\
    &\hspace{1em}- \left(\prod_{m = 0}^{n - 1} \mathcal D_{n - m}\right) \int_{u_0}^{u} \ud u_1\; \mathcal F_0(u_1)
  \end{split}
\end{equation}
and
\begin{equation} \label{eqn:shear_contribution mag}
  \begin{split}
    \frac{1}{4} &\left(\prod_{m = 0}^{n - 1} \mathcal D_{n - m}\right) \mathscr D_A \mathscr D_B [(^* C)^{AB} (u) - (^* C)^{AB} (u_0)] \\
    &= -\frac{\partial^n}{\partial u^n} \Delta \tilde Q_n^* (u, u_0) + \sum_{m = 0}^{n - 1} \left(\prod_{k = 0}^{m - 1} \mathcal D_{n - k}\right) \left(\frac{\ud}{\ud u}\right)^{n - m - 1} \left[{\mathcal F}_{n - m}^* (u) + {\mathcal G}_{n - m}^* (u)\right].
  \end{split}
\end{equation}
This allows us to identify a portion of the shear that gives rise to a particular portion of a moment of the news. 
The strain corresponding to the moment of the news is a time-dependent gravitational-wave signal that an interferometer could more easily measure.

Note that, due to the presence of the angular operators (with nontrivial kernel) on the left-hand sides of these equations, there is no way to recover certain multipoles of the shear.
By Eq.~\eqref{eqn:annihilator}, it follows that if one is considering this equation for some $n$, one can only recover multipoles with $l \geq n$.
This is discussed further in~\cite{Grant:2021hga, Freidel:2021ytz, Compere:2022zdz, Geiller:2024bgf}, and is related to the existence of generalizations of the Newman-Penrose constants which appear at linear order~\cite{Newman:1968uj}.
In~\cite{Compere:2022zdz}, the $l < n$ multipoles of the charges are called the ``memory-less'' parts of the charge, as they do not constrain the shear.

\section{Multipolar expansion of the fluxes' contributions to the moments} \label{sec:MultipoleFlux}

In this section, we present general expressions for the flux and pseudo-flux contributions to the zeroth, first, and second moments of the news, when the fields that enter into the fluxes are expanded in terms of multipole moments.
The ``scalarized'' (pseudo-) fluxes, charges and moments of the news used here, being all spin-weight-zero scalars, are expanded in terms of the basis of scalar spherical harmonics.
Specifically, for any scalar $S$, we write
\begin{equation}
    S = \sum_{l = 0}^\infty \sum_{m = -l}^l S_{lm} Y_{lm}.
\end{equation}
The shear, however, is a rank two STF tensor on the sphere, which can be expanded in terms of tensor harmonics (for more details, see~\cite{Thorne:1980ru}):
\begin{equation} \label{eqn: shear TH exp}
    C_{AB} = \sum_{l=2}^{\infty} \sum_{m=-l}^{l}\; \sum_{\mathrm I\in \{\mathrm E, \mathrm B\}} C^{\mathrm I}_{lm} (T^{\mathrm I}_{lm})_{AB}.
\end{equation}
Here, the coefficients $C^{\mathrm I}_{lm}$ (i.e., $C^{\mathrm E}_{lm}$ and $C^{\mathrm B}_{lm}$) are functions of retarded time, and $(T^{\mathrm I}_{lm})_{AB}$ [i.e., $(T^{\mathrm E}_{lm})_{AB}$ and $(T^{\mathrm B}_{lm})_{AB}$] are tensors on the 2-sphere with either electric ($\mathrm I = \mathrm E$) or magnetic ($\mathrm I = \mathrm B$) parity.
We review definitions and properties of these tensor harmonics in~\ref{app:STFtensorIdentities}.

We first summarize the results for the multipolar expressions for the zeroth and first fluxes, for which identical or equivalent formulas have been calculated previously.
The results will be useful for defining our notation and discussing the method of calculation.
We then give what we believe to be new expressions for the multipolar expansion of the second fluxes and pseudo-fluxes.
However, there have been closely related calculations of the Bondi charges that contain nonlinear terms~\cite{Blanchet:2023pce}, though involving only up to the quadrupole moment (and no higher multipoles).

\subsection{Zeroth and first fluxes}

The zeroth flux can be computed by taking the $u$ derivative of the expression~\eqref{eqn: shear TH exp} to obtain the multipolar expansion of the news tensor.
This multipolar expansion can then be substituted into Eq.~\eqref{eqn: F0}.
After re-writing the electric- and magnetic-parity tensor harmonics in terms of spin-weighted harmonics using Eqs.~\eqref{eqn: Tensor har exp}, the scalar multipolar moment $(\mathcal{F}_0)_{lm}$ of the flux can be written in terms of the integral of a product of three spin-weighted spherical harmonics.
These integrals can be written as proportional to the product of two Clebsch-Gordan coefficients with several selection rules determining when these integrals are nonzero.
These expressions arise because the product of two spin-weighted spherical harmonics can be written as a sum of spin-weighted spherical harmonics multiplied by products of Clebsch-Gordan coefficients.
The relevant expression is given in Eq.~\eqref{eq:sYlmProduct}.
The result has been derived before in several different notations (e.g.,~\cite{Favata:2008yd,Nichols:2017rqr,Mitman:2020pbt}); our result is most similar to the expression for $\Delta\Phi_{lm}$ in~\cite{Nichols:2017rqr}, though we use the slightly more compact notation that we introduce below.

In our notation, this expression is given by
\begin{align} \label{eqn:zeroth moment exp}
    (\mathcal{F}_0)_{lm}=& -\frac{1}{16} \sum_{\mathrm I', \mathrm I''} \sum_{l', m'} \sum_{l'', m''} \eta^{\mathrm E \mathrm I' \mathrm I''}_{l l' l''} \dot{C}^{\mathrm I'} _{l'm'} \dot{C}^{\mathrm I''} _{l''m''} \mathscr C^{lm2}_{l' m' l'' m''}. 
\end{align}
We suppress the range of the summations for ease of notation, but the relevant ranges are as follows: 
\begin{align}\label{eqn: range of summation}
 l', l'' \geq 2, \quad -l'\leq m'\leq l' \quad, -l''\leq m''\leq l'', \quad \text{and} \quad \mathrm I \in \{\mathrm E, \mathrm B\}. 
\end{align}
The coefficient $\mathscr C^{lm2}_{l' m' l'' m''}$ that appears in Eq.~\eqref{eqn:zeroth moment exp} is the $s=2$ case of a coefficient $\mathscr C^{lms}_{l' m' l'' m''}$, which can be written in terms of Clebsch-Gordan coefficients as
\begin{equation} \label{eqn:Def Coeff C}
    \mathscr C^{lms}_{l' m' l'' m''} = (-1)^{l+l'+l''}\sqrt{\frac{(2l'+1)(2l''+1)}{4\pi(2l+1)}} \langle l',-s,l'',s;l,0\rangle \langle l',m',l'',m'';l,m\rangle. 
\end{equation}
These coefficients are nonzero for 
\begin{align}\label{eqn: domain of def of coefficient C}
   \max(|l'-l''|,|m'+m''|,|s'+s''|)\leq l \leq l'+l'', \quad m=m'+m'' \quad \text{and} \quad  -l\leq m \leq l . 
\end{align}
Finally, we have introduced the coefficient $ \eta^{\mathrm I \mathrm I' \mathrm I''}_{ll'l''}$, which is given by 
\begin{align} \label{eq:etaDef}
    \eta^{\mathrm I \mathrm I' \mathrm I''}_{l l' l''}=
    \begin{cases}
        [1+(-1)^{l + l' + l''}]\hspace{2em} & \mathrm I = \mathrm E, \ \mathrm I' = \mathrm I'' = \mathrm E, \mathrm B \\
        i[1-(-1)^{l + l' + l''}]\hspace{2em} & \mathrm I = \mathrm E, \ \mathrm I' = \mathrm E, \ \mathrm I'' = \mathrm B \\
        -i[1-(-1)^{l + l' + l''}]\hspace{2em} & \mathrm I = \mathrm E, \ \mathrm I' = \mathrm B, \ \mathrm I'' = \mathrm E \\
        i[1-(-1)^{l + l' + l''}]\hspace{2em} & \mathrm I = \mathrm B, \ \mathrm I' = \mathrm I'' = \mathrm E, \mathrm B \\
        -[1+(-1)^{l + l' + l''}]\hspace{2em} & \mathrm I = \mathrm B, \ \mathrm I' = \mathrm E, \ \mathrm I'' = \mathrm B \\
        [1+(-1)^{l + l'+ l''}]\hspace{2em} & \mathrm I = \mathrm B, \ \mathrm I' = \mathrm B, \ \mathrm I'' = \mathrm E
    \end{cases}.
\end{align}
We will use these quantities in the other multipole expansions of the fluxes and pseudo-fluxes, which is why we have introduced expressions with $\mathrm I = \mathrm B$ that are not used in Eq.~\eqref{eqn:zeroth moment exp}.
These coefficients arise from the expansion of a product of two tensor harmonics; for more details, see~\ref{app:STFtensorIdentities}.

With the notation set, we can describe more briefly how we obtain the multipolar expressions for the first fluxes $\mathcal F_1$ and $\mathcal F_1^*$. 
Starting from Eqs.~\eqref{eqn:F1 elec} and~\eqref{eqn:F1 mag}, we then substitute in the multipolar expansion of $C_{AB}$ (and its $u$ derivative) in~\eqref{eqn: shear TH exp} and make use of the ``raising'' and ``lowering'' expressions for STF tensor harmonics in Eqs.~\eqref{eq:raise_T_EB} and~\eqref{eq:lower_T_EB}, respectively.
The result can be written as 
\begin{subequations}
    \label{eqn: First moment exp}
    \begin{align}
        (\mathcal{F}_1)_{lm} = & -\frac{1}{32} \sum_{\mathrm I', \mathrm I''} \sum_{l', m'} \sum_{l'', m''} \eta^{\mathrm E \mathrm I' \mathrm I''}_{l l' l''}[\dot{C}^{\mathrm I'} _{l'm'} {C}^{\mathrm I''} _{l''m''} - {C}^{\mathrm I'} _{l'm'} \dot{C}^{\mathrm I''} _{l''m''}]\mathscr B^{lm}_{l'm'l''m''}, \label{eqn:First moment flux exp elec} \\
        (\mathcal{F}_1 ^*)_{lm} = & -\frac{1}{32}\sum_{\mathrm I', \mathrm I''} \sum_{l',m'} \sum_{l'',m''} {\eta^{\mathrm B \mathrm I' \mathrm I''}_{l l' l''}}[\dot{C}^{\mathrm I'} _{l'm'} {C}^{\mathrm I''} _{l''m''} - {C}^{\mathrm I'} _{l'm'} \dot{C}^{\mathrm I''} _{l''m''}]\mathscr B^{lm}_{l'm'l''m''}. \label{eqn:First moment flux exp mag}
    \end{align}
\end{subequations}
The coefficients $\mathscr B^{lm}_{l'm'l''m''}$ are defined by
\begin{align} \label{eqn: Def coeff B}
    \mathscr B^{lm}_{l'm'l''m''} \equiv {} & \sqrt{(l''-2)(l'-2)(l''+3)(l'+3)} \mathscr C^{lm3}_{l' m' l'' m''}  \nonumber\\
    &+3\sqrt{(l'-1)(l''-1)(l'+2)(l''+2)} \mathscr C^{lm1}_{l' m' l'' m''} \nonumber\\
    & +[ (l''-2)(l''+3) +3(l'-1)(l'+2)]\mathscr C^{lm2}_{l' m' l'' m''}, 
\end{align}
and the sums run over the same ranges given in Eq.~\eqref{eqn: range of summation}.
Note that the expression for $(\mathcal{F}_1)_{lm}$ is consistent with the expression for the flux contribution to the CM memory in~\cite{Nichols:2018qac} and $(\mathcal{F}_1^*)_{lm}$ is consistent with the flux contribution to the spin memory in~\cite{Nichols:2017rqr}.
The different form of the coefficients in Eq.~\eqref{eqn: Def coeff B} from those in~\cite{Nichols:2017rqr,Nichols:2018qac} arises from how the integral was evaluated.
In~\cite{Nichols:2017rqr,Nichols:2018qac}, the flux used was a vector rather than a scalar, and a scalar integral was computed by contracting the vector flux with a vector spherical harmonic; here the fluxes are scalars and the multipoles are computed using an overlap integral with a scalar harmonic.
One can verify that these different expressions for the $\mathscr B^{lm}_{l'm'l''m''}$ coefficients are actually equivalent, after using identities involving the Clebsch-Gordan coefficients.
This should not be surprising, because the two expressions are equivalent after integrating by parts.

Note also that the electric and magnetic fluxes are related by the transformation $\eta^{\mathrm E \mathrm I' \mathrm I''}_{l l' l''} \leftrightarrow \eta^{\mathrm B \mathrm I' \mathrm I''}_{l l' l''}$.
Namely, writing $(\mathcal{F}_1)_{lm}$ as a function of $\eta^{\mathrm E \mathrm I' \mathrm I''}_{ll'l''}$, one can obtain $(\mathcal{F}^{*}_1)_{lm}$ via the relationship
\begin{equation}
    (\mathcal{F}^{*}_1)_{lm} = (\mathcal{F}_1)_{lm} (\eta^{\mathrm E \mathrm I' \mathrm I''}_{ll'l''} \rightarrow \eta^{\mathrm B \mathrm I' \mathrm I''}_{ll'l''}).
\end{equation}
We will find similar relationships for the second flux and pseudo-fluxes.

\subsection{Second flux and pseudo-fluxes}

Now we turn to computing the second flux and pseudo-fluxes.
Because there are relationships that transform the electric part of the fluxes into the magnetic-parity parts (and vice versa), we will write the expressions for the electric part only, and we give the transformation rules to obtain the magnetic parts at the end of the section.

We start with the electric-parity flux $\mathcal F_2$ in Eq.~\eqref{eqn: F2 elec}.
Because it is cubic in the shear and news, we can successively apply the product expression for two spin-weighted spherical harmonics in Eq.~\eqref{eq:sYlmProduct}.
The term of the form $N_{CD} C^{CD}$ in the flux has an expansion in spherical harmonics similar to that in the zeroth flux: i.e., 
\begin{equation} \label{eq:CabNab}
    C_{CD} N^{CD} = \frac{1}{2} \sum_{\mathrm I', \mathrm I''} \sum_{l',m'} \sum_{l'',m''} \sum_{\bar l, \bar m} \eta^{\mathrm E \mathrm I' \mathrm I''}_{l l' l''} \dot{C}^{\mathrm I'}_{l'm'} C^{\mathrm I''} _{l''m''} \mathscr C^{\bar l \bar m2}_{l' m' l'' m''} Y_{\bar l\bar m} ,
\end{equation}
where $\bar l$ and $\bar m$ are nonzero for the same ranges as $l$ and $m$ in Eq.~\eqref{eqn: domain of def of coefficient C} with $s'+s''=0$.

Using Eq.~\eqref{eq:CabNab} and several of the STF tensor-harmonic identities in~\ref{app:STFtensorIdentities}, the calculation of the spherical-harmonic modes of the flux $(\mathcal{F}_2)_{lm}$ again reduces to an integral over three spin-weighted spherical harmonics, though now with additional sums over $\bar l$ and $\bar m$ in the range in Eq.~\eqref{eqn: domain of def of coefficient C} with $s'+s''=0$ as well as $l'''$ and $m''$ over the analogous ranges as those in Eq.~\eqref{eqn: range of summation} for $l'$ and $m'$.
After a lengthy calculation, the result for the second flux is given by
\begin{align} \label{eqn:Second moment flux exp elec}
    (\mathcal{F}_2)_{lm} = \frac{1}{16} \sum_{\mathrm I', \mathrm I'', \mathrm I'''} \sum_{l',m'} \sum_{l'',m''} \sum_{\bar l,\bar m} \sum_{l''',m'''} &\eta^{\mathrm E \mathrm I' \mathrm I''}_{\bar l l' l''} \eta^{\mathrm E \mathrm E \mathrm I'''}_{l \bar l l'''} \dot{C}^{\mathrm I'} _{l'm'} {C}^{\mathrm I''} _{l''m''} {C}^{\mathrm I'''} _{l'''m'''} \mathscr C^{\bar l \bar m 2}_{l' m' l'' m''} \nonumber \\
    \times &\left[\sqrt{\frac{(l'''+2)!}{2(l'''-2)!}}\delta_{\mathrm E \mathrm I'''} \mathscr C^{lm 0}_{\bar l \tilde {m} l''' m'''}+\mathscr D^{lm} _{\bar l \bar m l'''m'''}\right], 
\end{align}
where the ranges of $\bar l$ and $\bar m$ have already been described, and where $l$ and $m$ should satisfy
\begin{equation}
   \max(|l'''-\bar{l}|,|m'''+\bar{m}|)\leq l \leq l'''+\bar{l}, \quad
   \text{and} \quad m=\bar{m}+m''',\\
\end{equation}
for the modes of the flux to be nonzero.
The new coefficient $\mathscr D^{lm} _{l'm'l''m''}$ is defined by
\begin{equation} \label{eq:DcoeffDef}
    \mathscr D^{lm} _{l'm'l''m''} \equiv 2\sqrt{\frac{{l'}({l'}+1)(l''+2)(l''-1)}{2}} \mathscr C^{lm1} _{l'm'l''m''}+\sqrt{\frac{({l'}+2)!}{2({l'}-2)!}} \mathscr C^{lm2} _{l',m',l'',m''} ,
\end{equation}
and $\delta_{\mathrm I \mathrm I'}$ is the Kronecker delta symbol (i.e., it vanishes unless $\mathrm I = \mathrm I'$, in which case it equals to $1$).

Next, we turn to the calculation of the second pseudo-fluxes.
These pseudo-fluxes are quadratic in the Bondi metric functions, which makes their calculation more similar to that of the zeroth and first fluxes.
Given the more complicated form of the angular operators acting on the shear in the radiative pseudo-flux in Eq.~\eqref{eqn: G rad2 elec}, the form of the spherical harmonic modes $(\mathcal{G}_2 ^{\mathrm{rad}})_{lm}$ are a bit lengthier, but the calculation of them proceeds in an analogous way to that of the first flux.
The result of the calculation for the electric part is 
\begin{align} \label{eqn:G rad exp elec}
    (\mathcal{G}_2 ^{\mathrm{rad}})_{lm}=& \frac{1}{16}\sum_{\mathrm I', \mathrm I''} \sum_{l',m'} \sum_{l'',m''} {\eta^{\mathrm E \mathrm I' \mathrm I''}_{l l' l''}} {C}^{\mathrm I'} _{l'm'} {C}^{\mathrm I''} _{l''m''}\biggr[\frac{1}{2}\sqrt{\frac{(l'+2)!\;(l''+2)!}{(l'-2)!\;(l''-2)!}} \delta_{\mathrm I' \mathrm B} \delta_{\mathrm I'' \mathrm B} \mathscr C^{lm0}_{l'm'l''m''} \nonumber \\ 
    &+ \delta_{\mathrm I' \mathrm B}\biggr(\sqrt{\frac{(l'+2)! \;l'(l'+1)(l''+2)(l''-1)}{(l'-2)!}}{\mathscr C^{lm1} _{l'm'l''m''}}+ \frac{(l'+2)!}{2(l'-2)!}\mathscr C^{lm2}_{l'm'l''m''}\biggr)\nonumber\\
    &+\frac{l(l+1)-2}{12} \mathscr B^l _{l'm'l''m''} \biggr].  
\end{align}
The nonradiative pseudo-flux in Eq.~\eqref{eqn: G nonrad2 elec} involves a product of the mass aspect and shear.
We therefore must expand the mass aspect in scalar (spin-weight zero) spherical harmonics and the shear in rank-2 STF tensor harmonics.
The calculation proceeds similarly to that of the other quadratic fluxes and pseudo-fluxes, and we find that  
\begin{align}\label{eqn:mC exp elec}
       (\mathcal{G}_2 ^{\mathrm{nonrad}})_{lm}=& \frac{1}{4}\sum_{l',m'}\sum_{l'',m''}\sum_{\mathrm I''} \eta^{\mathrm E \mathrm E \mathrm I''}_{ll'l''} m_{l'm'} C^{\mathrm I''}_{l''m''} \biggr[\sqrt{\frac{(l''+2)!}{2(l''-2)!}} \delta_{\mathrm E \mathrm I''} \mathscr C^{lm0}_{l'm'l''m''} \nonumber \\
       &+ \biggr(\sqrt{2l'(l'+1)(l''+2)(l''-1)}\mathscr C^{lm1}_{l'm'l''m''}+ \sqrt{\frac{(l'+2)!}{2(l'-2)!}}\mathscr C^{lm2}_{l'm'l''m''}\biggr)\biggr],
\end{align}
where now the range of $l'$ differs from other cases but the remaining sums run over equivalent ranges:
\begin{align}
    l'\geq0, \quad l'' \geq 2, \quad -l'\leq m'\leq l' \quad \text{and} \quad -l''\leq m''\leq l''.
\end{align}
The expression in Eq.~\eqref{eqn:mC exp elec} is fully general, but it involves the spherical-harmonic modes of the mass aspect, $m_{lm}$, which by the evolution equation for the Bondi mass aspect~\eqref{eqn:dot_m} have solutions that depend nonlinearly on the shear.
However, if one linearizes the evolution equation for the mass aspect, the time dependence of the modes $m_{lm}$ with $l \geq 2$ is then determined by that of the shear, and for $l < 2$, the mass aspect is constant.
Thus, for all $l$ we can write $m$ as
\begin{equation}
    m_{lm}= \mu_{lm}+ \frac{1}{4}\sqrt{\frac{(l+2)!}{2(l-2)!}} {C}^{\mathrm E}_{lm} + O(C^2) ,
\end{equation}
where the modes $\mu_{lm}$ with $l \geq 0$ are constant and the modes ${C}^{\mathrm E}_{lm}$ are the electric-parity multipole moments of the shear, which are nonzero for $l \geq 2$. 

In the post-Newtonian calculations that we perform below, it will be useful to specialize to systems that are stationary in the past ($u\rightarrow -\infty$) as is often assumed in the post-Newtonian context~\cite{Blanchet:2013haa}.
In this case, all moments of the initial mass aspect with $l>0$ must vanish~\cite{Flanagan:2015pxa}, so that $\mu_{00}$ is the only nonvanishing moment of $\mu_{lm}$.
In terms of the initial mass $M$ of the system,
\begin{equation}
    \mu_{00} = \sqrt{4\pi} M.
\end{equation}
Thus, with the additional approximation of past stationarity, Eq.~\eqref{eqn:mC exp elec} can be written as
\begin{align} \label{eqn:G nonrad exp elec}
    (\mathcal{G}_2 ^{\mathrm{nonrad}})_{lm} & = \frac{1}{2} M \sqrt{\frac{(l+2)!}{2(l-2)!}} {C}^{\mathrm E} _{lm} + \frac{1}{8} \sum_{\mathrm I''} \sum_{l',m'} \sum_{l'',m''} {\sqrt{\frac{(l'+2)!}{2(l'-2)!}}}{\eta^{\mathrm E \mathrm E \mathrm I''}_{l l' l''}} {C}^{\mathrm E} _{l'm'}{C}^{\mathrm I''}_{l''m''} \biggr[\nonumber\\
    &\biggr( \frac{1}{2}\sqrt{\frac{(l'+2)!}{2(l'-2)!}} \mathscr C^{lm2} _{l'm'l''m''} + \sqrt{\frac{l'(l'+1)(l''+2)(l''-1)}{2}} \mathscr C^{lm1} _{l'm'l''m''} \biggr) \nonumber\\
    &+ \frac{1}{2} \delta_{\mathrm E \mathrm I''} \sqrt{\frac{(l''+2)!}{2(l''-2)!}} \mathscr C^{lm0} _{l'm'l''m''}\biggr] + O(C^3) .
\end{align}
The sums over $l'$ and $l''$ are again restricted to $l',l''\geq 2$.
As we will discuss in more detail in Sec.~\ref{sec:Decompostion of PN expansion}, the term proportional to $M$ looks similar to the tail terms in PN theory, which arise at 1.5 PN order relative to the corresponding multipolar mode of the waveform.
However, unlike the tail terms in post-Newtonian calculations (see, e.g.,~\cite{Blanchet:1987wq,Blanchet:1992br}) that involve an integral of multipole moments that correspond to two derivatives of the strain times the mass monopole scaled by a logarithmic function, these tail-like terms from the pseudo-flux are instantaneous terms involving the product of the mass monopole and the news.
We discuss the discrepancy between the forms of these two terms in more detail in Sec.~\ref{sec:Decompostion of PN expansion} and~\ref{sec: PN order of Moments}.

The magnetic pieces can now be obtained through the following transformation rules applied to the expressions for the (pseudo-)fluxes, when they are considered to be functions of $\eta^{\mathrm E \mathrm I' \mathrm I''}_{ll'l''}$, $\delta_{\mathrm E \mathrm I''}$ and $\delta_{\mathrm I'' \mathrm B}$:
\begin{subequations}{\label{eqn:elec-mag flux transformation}}
\begin{align}
    (\mathcal{F}^{*}_2)_{lm} & = (\mathcal{F}_2)_{lm} (\eta^{\mathrm E \mathrm E \mathrm I'''}_{l\bar{l}l'''} \rightarrow \eta^{\mathrm B \mathrm E \mathrm I'''}_{l\bar{l}l'''} , \delta_{\mathrm E \mathrm I'''} \rightarrow \delta_{\mathrm B \mathrm I'''}), \\
    (\mathcal{G}_2 ^{\mathrm{rad}\;*})_{lm} & = (\mathcal{G}_2 ^{\mathrm{rad}})_{lm} (\eta^{\mathrm E \mathrm I' \mathrm I''}_{ll'l''} \rightarrow\eta^{\mathrm B \mathrm I' \mathrm I''}_{ll'l''}, \delta_{\mathrm I'' \mathrm B} \rightarrow \delta_{\mathrm I'' \mathrm E}) , \label{eqn:G2rad_transform} \\
    (\mathcal{G}_2 ^{\mathrm{nonrad}\;*})_{lm} &= (\mathcal{G}_2 ^{\mathrm{nonrad}})_{lm} (\eta^{\mathrm E \mathrm E \mathrm I''}_{l\bar{l}l''}\rightarrow \eta^{\mathrm B \mathrm E \mathrm I''}_{l\bar{l}l''}, \; \delta_{\mathrm E \mathrm I''}\rightarrow \delta_{\mathrm B \mathrm I''}),
\end{align}
\end{subequations}
where the last transformation applies to the original expression for the pseudo-flux in Eq.~\eqref{eqn:mC exp elec}, not the derived expression in Eq.~\eqref{eqn:G nonrad exp elec}.
Furthermore, these transformations should be applied only to the terms with the same dummy indices in the expressions for the (pseudo-)fluxes; for example, Eq.~\eqref{eqn:G2rad_transform} implies that the factors of $\delta_{\mathrm I' \mathrm B}$ in Eq.~\eqref{eqn:G rad exp elec} should \emph{not} be transformed.
Given these transformation rules, we do not need to explicitly provide the magnetic-parity analogs of Eqs.~\eqref{eqn:Second moment flux exp elec}, \eqref{eqn:G rad exp elec} and~\eqref{eqn:mC exp elec}.

\section{Post-Newtonian formalism and signals from compact binaries}\label{sec:Decompostion of PN expansion} 

In this section, we evaluate the PN expressions for the moments of the news and the corresponding parts of the shear for the zeroth, first and second fluxes as well as the second pseudo-fluxes.
We first review the relationship between the radiative data in the Bondi-Sachs framework with that in the post-Newtonian formalism.
We next compute the shear corresponding to the moments of the news (as defined in Sec.~\ref{subsec:shear-from-moments}).
Finally, we show, for the quadrupole and octupole, that the different flux contributions to the moments of the news can account for all of the PN memory terms.
Also for these modes, we show that the nonlinear, instantaneous terms that appear in the PN expression for the waveform at 3.5~PN and 3~PN order for the quadrupole and octupole, respectively, can also be obtained from the fluxes and pseudo-fluxes up to second total time-derivative terms.
These remainder terms should, in principle, arise from charge contributions or higher flux contributions that we do not compute.

\subsection{Relation between post-Newtonian and Bondi-Sachs waveforms}

To perform post-Newtonian calculations using our results in the Bondi formalism, we need to relate gravitational-wave data in the PN formalism to that in the Bondi-Sachs formalism.
As shown by Blanchet et al.\ in~\cite{Blanchet:2020ngx} (where a coordinate transformation is derived that maps the metric in harmonic gauge to that in the Bondi gauge), the Bondi shear is related to the gravitational-wave strain $H_{ij}^\mathrm{TT}$ in transverse traceless gauge by
\begin{equation} \label{eqn:shear leading order}
    H_{ij}^{TT} e^i{}_A e^j{}_B = \frac{1}{r} C_{AB} + O(1/r^2). 
\end{equation}
The quantities $e^i{}_{A}$ are given by $e^i{}_A = \partial n^i/\partial \theta ^A$, where $n^i = x^i/r$.
The Cartesian coordinates $x^i$ are defined from specializing the arbitrary angular coordinates $\theta^A$ on the two-sphere to spherical polar coordinates, and constructing Cartesian coordinates from $r$ and the other two angular coordinates using the standard flat-space relationships. 
The strain on the left-hand side is typically computed in a set of radiative coordinates which are consistent with Penrose's notion of asymptotic simplicity or flatness~\cite{Penrose:1962ij} (see Sec.~2.1.5 of~\cite{Blanchet:2013haa} for further details\footnote{Section and equation numbers here refer to the latest version (v5) of Ref.~\cite{Blanchet:2013haa} on the arXiv, which is organized differently from both earlier versions of the paper on arXiv and the current version published in \textit{Living Reviews in Relativity}.}).
Because $H_{ij}^{TT}$ is symmetric, transverse and traceless, contracting $e^i{}_A e^j{}_B$ into $H_{ij}^{TT}$ produces a tensor that is STF in Bondi coordinates.

The strain can then be expanded in terms of a set of radiative multipole moments using tensor harmonics (see, e.g.,~\cite{Thorne:1980ru}):
\begin{equation} \label{eq:HijTT-UV}
    H_{ij}^{TT} = \frac{1}{r} \sum_{l,m} [U_{lm} (T^{\mathrm E}_{lm})_{ij} + V_{lm} (T^{\mathrm B}_{lm})_{ij}].
\end{equation}
The tensor harmonics $(T^{\mathrm I}_{lm})_{ij}$ are symmetric, transverse-traceless tensors, like $H_{ij}^{TT}$.
Nominally, they are a distinct set of tensor harmonics from those that we used in the context of the Bondi-Sachs framework, but we use a similar notation to describe them, because they are related to these other tensor harmonics by contracting with ${e^i}_A$ as follows:
\begin{equation}
    (T^{\mathrm I}_{lm})_{ij} e^i{}_A e^j{}_B = (T^{\mathrm I}_{lm})_{AB} .
\end{equation}
It is not necessary to take the STF part on the 2-sphere indices for the same reasons discussed below Eq.~\eqref{eqn:shear leading order}.
Comparing Eqs.~\eqref{eqn: shear TH exp} and~\eqref{eq:HijTT-UV}, it therefore follows that
\begin{equation} \label{eqn:shear-rad mode}
    C^{\mathrm E} _{lm} = U_{lm}, \qquad C^{\mathrm B} _{lm} = V_{lm} .
\end{equation}
One can also compute the polarizations $h_+$ and $h_\times$ via the relationship
\begin{equation} \label{eq:h-hplus-hcross}
    h  \equiv h_+ - i h_\times = \sum_{l,m} h_{lm} (_{-2}Y_{lm}) ,
\end{equation}
where the moments $h_{lm}$ are related to $U_{lm}$ and $V_{lm}$ by
\begin{equation} \label{eq:hlm-Ulm-Vlm}
    h_{lm} = \frac{1}{r\sqrt{2}}(U_{lm}-i V_{lm}) ,
\end{equation}
and where $_{-2}Y_{lm}$ is a spin-weighted spherical harmonic.
The conventions that we use for these harmonics are defined in~\ref{app:STFtensorIdentities}.

When comparing with PN results, we will also use the symmetric, trace-free $l$-index Cartesian multipole tensors.
We use the multi-index notation for representing STF tensor of rank $l$:
\begin{align}
    T_L= T_{\langle i_1 \cdots i_{l} \rangle}, \qquad n^L= n^{\langle i_1} \cdots n^{i_l \rangle},
\end{align}
where angle brackets around three-dimensional indices indicate to take the STF part with respect to the Cartesian metric $\delta_{ij}$.
There then exist a set of STF basis tensors $\mathcal Y^{lm}_L$ that relate the $U_{lm}$ and $V_{lm}$ modes to STF rank-$l$ tensors $\mathcal U_L$ and $\mathcal V_L$.
The relations are given by (see, e.g.,~\cite{Thorne:1980ru})
\begin{equation} \label{eqn:radiative moments}
    \mathcal{U}_L = a_l \sum_{m=-l}^l U_{lm} \mathcal{Y}^{lm} _L, \qquad 
    \mathcal{V}_L = - b_l \sum_{m=-l}^l V_{lm} \mathcal{Y}^{lm} _L,
\end{equation}
where we have defined
\begin{equation} \label{eq:a_l-b_l-def}
    a_l = \frac{l!}{4}\sqrt{\frac{2l(l-1)}{(l+1)(l+2)}} , \qquad
    b_l = \frac{(l+1)!}{8}\sqrt{\frac{2l(l-1)}{(l+1)(l+2)}} .
\end{equation}
Although we do not give the explicit expressions for the tensors $\mathcal Y^{lm} _{L}$ (see~\cite{Thorne:1980ru} instead), they are related to the scalar spherical harmonics $Y_{lm}$ via
\begin{align}
    \mathcal{Y}^{lm} _L n^L= Y_{lm}
\end{align}
(see also~\cite{Thorne:1980ru}).
Thus contracting Eq.~\eqref{eqn:radiative moments} with $n^L$ gives 
\begin{equation} \label{eqn: multi moments contracted} 
    \mathcal{U}_{L} n^L = a_l \sum_{m=-l}^l U_{lm} Y_{lm}, \qquad
    \mathcal{V}_L n^L = - b_l\sum_{m=-l}^l V_{lm} Y_{lm}. 
\end{equation}
The orthogonality of the $\mathcal Y^{lm} _{L}$ with different $l$ and $m$ allows the $U_{lm}$ and $V_{lm}$ moments to be determined from $\mathcal U_L$ and $\mathcal V_L$, respectively.
We will use these relations in the calculations in the next section, so we give them here (see~\cite{Thorne:1980ru,Favata:2008yd}):
\begin{equation} \label{eqn: mode-multi mom relation }
    U_{lm}= \frac{16 \pi }{(2 l+1)!!}\sqrt{\frac{(l+1) (l+2)}{2 l (l-1)}} \mathcal{U}_{L}\bar{\mathcal{Y}}^{lm} _{L}, \quad 
    V_{lm}=-\frac{32 \pi  l}{(2 l+1)!!} \sqrt{\frac{l+2}{2 l (l+1) (l-1)}} \mathcal{V}_{L} \bar{\mathcal{Y}}^{lm} _{L} .
\end{equation}
We use an overline to denote complex conjugation; the double factorial, $(2l+1)!!$, is the product of all odd integers less than or equal to $2l+1$.
To leading order in the multipolar post-Minkowski expansion, the radiative moments for a given $l$ are equal to $l$ time derivatives of a set of so-called ``canonical multipoles'':
\begin{equation} \label{eqn:radiative-source moment}
    \mathcal{U}_L = M_L ^{(l)} + O(G), \qquad \mathcal{V}_L= S_L ^{(l)} + O(G). 
\end{equation}
The tensor $M_L$ is the mass multipole moment and $S_L$ is the current multipole moment; the superscript $(l)$ indicates the number of derivatives with respect to retarded time. 
To compare our expressions with existing post-Newtonian results, we will use these canonical multipoles, as these multipoles are commonly used in this context.

\subsection{Post-Newtonian expressions for the shear associated with moments of the news} \label{subsec:PNmomentsShear}

Given the simple leading PN relationship between the multipoles of the Bondi shear and the canonical moments in Eqs.~\eqref{eqn:shear-rad mode}, \eqref{eqn: mode-multi mom relation } and~\eqref{eqn:radiative-source moment}, we can now compute the leading PN expressions for the multipolar expansions of the fluxes in Sec.~\ref{sec:MultipoleFlux} and the corresponding contributions to the waveforms associated with these fluxes.
Because each retarded-time derivative in Eq.~\eqref{eqn:radiative-source moment} is accompanied by a factor of $1/c$, higher $l$ multipoles are of a higher PN order.
This allows us to compute 3.5~PN-accurate quadrupole and octupole modes of the waveform that come from the (pseudo-)flux contributions to the zeroth, first and second moments of the news.
This calculation uses as input the leading, linearized quadrupole, octupole and hexadecapole expressions for the radiative or canonical moments.

We focus on these parts, in particular, because they will allow us to compare these contributions of the moments of the news to the shear with the corresponding PN expression for the radiative multipole moments in terms of the canonical moments (which appear, for example, in Sec.~2.4.5 of~\cite{Blanchet:2013haa}). 
For brevity, we will present the results for the purely electric contributions (i.e., $\mathcal U_L$); these results could be extended to the magnetic parts (the $\mathcal V_L$) using the methods that we use for the electric part.
We also give, for simplicity, just the quadrupole and octupole terms.
There are also contributions to the $l=4$ and $l=5$ moments, which occur at the same PN order as those we compute for the quadrupole and octupole.
We are able to show that all of the nonlinear ``memory'' terms (in the PN usage of ``memory'') can be determined from the flux contribution to the zeroth moment of the news.
The nonlinear instantaneous terms for the mass quadrupole and octupole moments can be written in terms of the flux terms that contribute to the first and second moments, up to total second derivatives.
Terms involving second derivatives could be calculated from (pseudo-)flux contributions from the third moment of the news, or from charge contributions, neither of which we compute in this paper.

In the remainder of this section, we present the results for the leading PN expressions for the shear associated with the moments of the news in the quadrupole and octupole modes.
We will use several of these results in Sec.~\ref{sec: PN order of Moments} when we specialize to compact-binary sources (though we will compute additional modes there).
We then describe our procedure to compare the PN expressions for the multipole moments of the strain (as given in Sec.~2.4.5 of~\cite{Blanchet:2013haa}) with the expressions that we derive from taking derivatives of moments of the news.

\subsubsection{Expressions for the quadrupole and octupole waveforms generated by fluxes and pseudo-fluxes}

To compute the contributions to the shear coming from the flux and pseudo-flux terms, we return to the expressions for the electric and magnetic parts of the shear in Eqs.~\eqref{eqn:shear_contribution elec} and~\eqref{eqn:shear_contribution mag} without the contributions from the changes in the charges.
We will specialize our computation to the contributions to the shear that come from moments less than the third; thus, we consider these equations with $n = 2$ and neglect the charge contributions.

Because there are angular operators acting on the shear (on the left-hand side) and fluxes and pseudo-fluxes (on the right-hand side), it is convenient to expand the left-hand side in spherical harmonics (in addition to the right-hand side, which we already expanded in Sec.~\ref{sec:MultipoleFlux}).
For each spherical harmonic mode, inverting the angular operators to determine the moments of the shear then becomes an algebraic operation.
Specifically, if we use $U^{\mathcal{F}}_{lm}$ and $V^{\mathcal{F}}_{lm}$ to denote the electric and magnetic parts of the shear that arise from the flux and pseudo-flux terms for a given spherical harmonic mode, then we can write the contribution from the zeroth, first and second moments relatively concisely.
To do so, we apply Eq.~\eqref{eq:lower_T_EB} twice to the expansion of $C_{AB}$ in terms of tensor harmonics and use the facts that $\mathcal D_1 = \mathscr D^2$, $\mathcal D_2 = \mathscr D^2+2$ and $\mathscr D^2 Y_{lm} = -l(l+1) Y_{lm}$.
It then follows that the electric part of the shear, $U^{\mathcal{F}}_{lm}$, is given by
\begin{align} \label{eqn: flux contribution}
    \frac 14 \sqrt{\frac{(l+2)!}{2(l-2)!}} U^{\mathcal{F}}_{lm} = & -\int_{-\infty}^u \ud u_1 (\mathcal{F}_0)_{lm}(u_1) + \frac{(l-1)!}{(l+1)!} (\mathcal{F}_1)_{lm} \nonumber \\
    &  -6 \frac{(l-2)!}{(l+2)!} \frac{\ud }{\ud u} \left[(\mathcal{F}_2)_{lm} + (\mathcal{G}_2 ^{\mathrm{rad}})_{lm} + (\mathcal{G}_2 ^{\mathrm{nonrad}})_{lm} \right] + \frac{\ud^2}{\ud u^2} (\cdots).
\end{align}
The product of the angular operators has the property that it annihilates spherical harmonics with $l \leq 1$ for the zeroth, first and second moments, and $l \leq n-1$ for the n\textit{th} moment.
Equation~\eqref{eqn: flux contribution} holds for $l\geq 2$ and for the corresponding permitted $m$.
The ellipsis at the end of Eq.~\eqref{eqn: flux contribution} indicates that we neglected contributions from the fluxes and pseudo-fluxes that contribute to higher moments of the news (starting at the third moment); however, for $l=2$, the angular operators set the contribution from the third flux to zero (so no additional flux terms contribute).
We are also neglecting all charge terms, which is why we add the superscript $\mathcal F$ to $U^{\mathcal{F}}_{lm}$.
Finally, the magnetic part $V^{\mathcal{F}}_{lm}$ can be obtained by substituting the magnetic fluxes and pseudo-fluxes on the right-hand side of Eq.~\eqref{eqn: flux contribution}, and using the fact that $\mathcal F^*_0 = 0$.

The expression in~\eqref{eqn: flux contribution} for the shear related to the zeroth, first and second moments of the news holds to any PN order, when the full expressions for the multipolar expansion of the fluxes in Sec.~\ref{sec:MultipoleFlux} are used.
To compare with existing PN results and to compute the shear from specific astrophysical sources, we will substitute into our expressions for the fluxes the necessary linearized PN radiative or canonical moments.
In addition, we will give explicit expressions for the mass quadrupole and octupole moments (namely, $U^{\mathcal{F}}_{2m}$ and $U^{\mathcal{F}}_{3m}$). 
These moments up to 3.5~PN and 3~PN orders for the quadrupole and octupole, respectively, are summarized in Sec.~2.4.5 of~\cite{Blanchet:2013haa}.
We begin by writing the expressions in terms of the radiative moments.

The expressions for $U^{\mathcal{F}}_{3m}$ are simpler, because they can be computed from the leading (linearized) PN parts of the radiative quadrupole and octupole moments.
The zeroth and first flux, as well as the second radiative pseudo-flux contributions to $U^{\mathcal{F}}_{3m}$, all arise at 2.5~PN order relative to the leading octupole waveform (which is 0.5~PN orders higher than the leading quadrupole waveform).
The nonradiative pseudo-flux gives rise to a term at relative 1.5~PN order in addition to the relative 2.5~PN-order contribution from the other fluxes and pseudo-fluxes.
The second flux contribution to the strain first enters at greater than 4~PN order, which is the highest-order waveform computed to date~\cite{Blanchet:2023sbv}; thus, we will not compute the flux part of the second moment.
Thus, the terms in Eq.~\eqref{eqn: flux contribution}, which contribute to the leading and next-to-leading order parts of $U^{\mathcal F}_{3m}$, are the following:
\begin{subequations} \label{eqn:flux octu expansion}
    \begin{align} 
        (\mathcal{F}_{0})_{3m} = & -\frac{1}{4} \biggr(\frac{G}{c^5}\biggr) \sum_{m',m''}(\dot{U} _{2m'} \dot{U}_{3m''}\mathscr C^{3m2}_{2m'3m''} +i\dot{U} _{2m'} \dot{V} _{2m''}\mathscr C^{3m2}_{2m'2m''}), \\
        (\mathcal{F}_{1})_{3m}= & -\frac{1}{16} \biggr(\frac{G}{c^5}\biggr) \sum_{m',m''} (\dot{U} _{2m'} {U} _{3m''}  - {U} _{2m'} \dot{U} _{3m''})(\mathscr B^{3m}_{2m'3m''}-\mathscr B^{3m}_{3m'2m''}) , \\
        (\mathcal{G}_{2}^{\mathrm{rad}})_{3m} = {} & \biggr(\frac{G}{c^5}\biggr)  \sum_{m',m''} \left[ 5{U} _{2m'} {U} _{3m''} \left(\mathscr C^{3m2}_{2m'3m''}+\frac{5}{4}\sqrt{10}\mathscr C^{3m1}_{2m'3m''}\right) \right. \nonumber \\
        &\left. +i{U}_{2m'} {V} _{2m''} \left(4\mathscr C^{3m1}_{2m'2m''}+\frac{11}{2}\mathscr C^{3m2}_{2m'2m''}\right)\right] , \\
        (\mathcal{G}_{2}^{\mathrm{nonrad}})_{3m} = {} & \biggr(\frac{G}{c^3}\biggr)  \sqrt{15} M U_{3m} + \frac{3}{2} \biggr(\frac{G}{c^5}\biggr) \sum_{m',m''} \big[i{U}_{2m'} {V} _{2m''}(\mathscr C^{3m2}_{2m'2m''}+2\mathscr C^{3m1}_{2m'2m''}) \nonumber \\
        &+ {U} _{2m'} {U} _{3m''} (6\mathscr C^{3m2}_{2m'3m''}+ 3\sqrt{10}\mathscr C^{3m1}_{2m'3m''}  +2\sqrt{5}\mathscr C^{3m0}_{2m'3m''}) \big] .
    \end{align}
\end{subequations}

We now compute the contributions to $U^{\mathcal{F}}_{2m}$.
For simplicity, we restrict to the expression known to 3.5~PN order (as in Sec.~3.3 of~\cite{Blanchet:2013haa}), rather than the new 4~PN result in~\cite{Blanchet:2023sbv}.
This computation will require us to compute the second radiative and nonradiative pseudo-fluxes, as well as the zeroth and first fluxes.
We can again neglect the second flux; its contribution enters at 5~PN order, which is one order above the known 4~PN result.
The expressions for these fluxes and pseudo-fluxes are somewhat lengthier than those of the octupole, and we present them in sequence below.

First, the contributions from $\mathcal{F}_0$ to the quadrupole moment can be obtained from the integral of the following multipole moments:
\begin{subequations} \label{eqn:flux-quad}
\begin{align}
  (\mathcal{F}_{0})_{2m} = & -\frac{1}{8} \biggr(\frac{G}{c^5}\biggr) \sum_{m',m''} \dot{U} _{2m'} \dot{U}_{2m''}\mathscr C^{2m2}_{2m'2m''} \nonumber\\
  & -\frac{1}{8} \biggr(\frac{G}{c^7}\biggr) \sum_{m',m''}(2\dot{U} _{2m'} \dot{U}_{4m''}\mathscr C^{2m2}_{2m'4m''} + \dot{U} _{3m'} \dot{U}_{3m''}\mathscr C^{2m2}_{3m'3m''}\nonumber\\ 
  &+ \dot{V} _{2m'} \dot{V}_{2m''}\mathscr C^{2m2}_{2m'2m''} + 2i\dot{U} _{3m'} \dot{V} _{2m''}\mathscr C^{2m2}_{3m'2m''}+2i\dot{V} _{3m''} \dot{U}_{2m'} \mathscr C^{2m2}_{2m'3m''}) . 
\end{align}
Next, the contribution from $\mathcal{F}_1$ to the mass quadrupole is obtained from
\begin{align}
    (\mathcal{F}_{1})_{2m} = & -\frac{1}{16} \biggr(\frac{G}{c^7}\biggr) \sum_{m',m''} \left[(\dot{U} _{2m'} {U} _{4m''}  - {U} _{2m'} \dot{U} _{4m''})(\mathscr B^{2m}_{2m'4m''}-\mathscr B^{2m}_{4m'2m''}) \right.\nonumber\\ 
    & \left. + i (\dot{U} _{3m'} {V} _{2m''}  - {U} _{3m'} \dot{V} _{2m''} + \dot{V} _{3m'} {U} _{2m''}  - {V} _{3m'} \dot{U} _{2m''}) \mathscr B^{2m}_{3m'2m''} \right] ,
\end{align}
up to a numerical factor.
Finally for the second moment, the 2.5 and 3.5~PN order contributions come from the pseudo-flux terms, and they are given by the derivatives of
\begin{align}
    (\mathcal{G}_{2}^{\mathrm{rad}})_{2m} = {} & \frac{1}{2} \biggr(\frac{G}{c^5}\biggr) \sum_{m',m''} {U} _{2m'} {U} _{2m''} (\mathscr C^{2m2}_{2m'2m''}+\mathscr C^{2m1}_{2m'2m''}) \nonumber\\
    &+ \frac{1}{2} \biggr(\frac{G}{c^7}\biggr) \sum_{m',m''} \biggr[ {U} _{3m'} {U} _{3m''} (4\mathscr C^{2m3}_{3m'3m''}+ 6 \mathscr C^{2m2}_{3m'3m''}+ 5 \mathscr C^{2m1}_{3m'3m''}) \nonumber\\ 
    & + \frac 13 {U} _{2m'} {U} _{4m''} (20 \mathscr C^{2m2}_{2m'4m''}+ 9\sqrt{2}\mathscr C^{2m1}_{2m'4m''}) \nonumber \\
    & + {V}_{2m'} {V} _{2m''}(4\mathscr C^{2m2}_{2m'2m''}+ 7\mathscr C^{2m1}_{2m'2m''}+3\mathscr C^{2m0}_{2m'2m''}) \nonumber\\
    &+i {U} _{3m'} {V} _{2m''} (7\mathscr C^{2m2}_{3m'2m''}+4\sqrt{10}\mathscr C^{2m1}_{3m'2m''}) \nonumber\\
    &- i {V} _{3m'} {U} _{2m''}(7\sqrt{10}\mathscr C^{2m1}_{3m'2m''}+ 19 C^{2m2}_{3m'2m''}) \biggr] , \\
    (\mathcal{G}_{2}^{\mathrm{nonrad}})_{2m} = {} & \biggr(\frac{G}{c^3}\biggr) \sqrt{3} M U_{2m} + \frac 32 \biggr(\frac{G}{c^5}\biggr)  \sum_{m',m''} {U} _{2m'} {U} _{2m''} ( \mathscr C^{2m2}_{2m'2m''}+2\mathscr C^{2m1}_{2m'2m''}+  \mathscr C^{2m0}_{2m'2m''})  \nonumber \\
    &+ \frac 32 \biggr(\frac{G}{c^7}\biggr) \sum_{m',m''} \big[ 5{U} _{3m'} {U} _{3m''}(\mathscr C^{2m2}_{3m'3m''}+ 2\mathscr C^{2m1}_{3m'3m''} +\mathscr C^{2m0}_{3m'3m''} )\nonumber\\
    &  + {U} _{2m'} {U} _{4m''}(16\mathscr C^{2m2}_{2m'4m''} +13\sqrt{2}\mathscr C^{2m1}_{2m'4m''} + 2\sqrt{15}\mathscr C^{2m0}_{2m'4m''}) \nonumber \\ 
    &+ i  {U} _{3m'} {V} _{2m''}(5\mathscr C^{2m2}_{3m'2m''}+ 2\sqrt{10}\mathscr C^{2m1}_{3m'2m''}) \nonumber \\ 
    &- i{V} _{3m'} {U} _{2m''}(\mathscr C^{2m2}_{3m'2m''}+ \sqrt{10}\mathscr C^{2m1}_{3m'2m''})  \big].
\end{align}    
\end{subequations}
We will make use of these expressions in Sec.~\ref{sec: PN order of Moments}, when we compute the results from compact binary sources.
However, we will also need several other multipoles of the fluxes and pseudo-fluxes that we do not give in this section.

\subsubsection{Comparison with PN waveform multipoles}

The expressions for the multipoles $U^{\mathcal{F}}_{2m}$ and $U^{\mathcal{F}}_{3m}$ are convenient for some calculations, but in many post-Newtonian calculations, the STF rank-$l$ tensor multipoles $\mathcal U_L$ (specifically, in this case, $\mathcal U_{ij}$ and $\mathcal U_{ijk}$) are used instead.
Moreover, the PN expressions for these radiative multipoles are often written in terms of the canonical multipoles of Eq.~\eqref{eqn:radiative-source moment} instead of the radiative multipoles.
Comparing the PN waveform modes derived from a harmonic-gauge calculation with our waveform modes $U^{\mathcal{F}}_{lm}$ requires a nontrivial calculation. 

This is further complicated by the fact that several of the nonlinear terms in the PN harmonic-gauge modes are written in terms of canonical multipoles that correspond via Eq.~\eqref{eqn:radiative-source moment} to integrals of the radiative multipoles.
To compare our modes with the corresponding PN modes (in~\cite{Blanchet:2013haa}, for example), we must first use the product rule on these nonlinear terms to write it in a form closer to what we obtain from taking the appropriate number of time derivatives of each moment of the news.
To do this, we first collect the ``hereditary'' (also called ``tail'' and ``memory'') terms that involve retarded-time integrals of the waveform modes in the harmonic-gauge PN expressions; these terms do not need to be recast.
The remaining nonlinear terms are the so-called ``instantaneous'' terms.
We then manipulate these instantaneous terms with the product rule to write them as products of canonical multipoles, total first derivatives of these products and total second derivatives of these products.
For the product and first derivative of these products, we keep terms where the total number of derivatives acting on each individual canonical moment is either zero or one derivative of the corresponding radiative moment.
This still is not a unique prescription, in some cases, so we also require that the coefficient multiplying the product of moments of the same type (e.g., mass quadrupole times mass octupole) with different numbers of derivatives acting on the corresponding moment should have the same coefficient, up to a sign. 

The result of this lengthy rewriting on the harmonic-gauge PN waveform is given in~\ref{sec: Appendix 1}, with Eq.~\eqref{eqn:quadrupole PN exp decomposed} being the expression for the quadrupole and Eq.~\eqref{eqn:octupole PN exp decomposed} being the expression for the octupole.
This rewriting of the typical PN expressions puts them in a form more similar to Eq.~\eqref{eqn: flux contribution} (aside from the difference in the types of multipole moments used in~\ref{sec: Appendix 1}). 
We will now be able to determine that the terms in Eqs.~\eqref{eqn:quadrupole PN exp decomposed} and~\eqref{eqn:octupole PN exp decomposed} that are integrals of products of canonical moments correspond to the integral of the flux $\mathcal{F}_0$, the term $\mathcal{F}_1$ corresponds to the terms with no retarded time derivatives, and $\mathcal{F}_2$ is responsible for producing the term with the total first derivative.
For the octupole (but not the quadrupole), the total second-derivative terms could be written in terms of flux or pseudo-flux terms from the third moment; there will also be charge contributions to the PN expressions for both multipoles.
Although we have neglected the charge contributions, they would need to account for all remaining differences if the Bondi-Sachs and post-Newtonian calculations are consistent.

With the PN quadrupole $\mathcal U_{ij}$ and octupole $\mathcal U_{ijk}$ written in the form in Eqs.~\eqref{eqn:quadrupole PN exp decomposed} and~\eqref{eqn:octupole PN exp decomposed}, we can now more easily compare the different flux and pseudo-flux contributions in $U^{\mathcal{F}}_{2m}$ and $U^{\mathcal{F}}_{3m}$, which we can do separately for each flux.
What now remains is to convert between our expressions involving scalar multipoles and the STF tensor multipoles used in PN theory.
This will require evaluating the combination of Clebsch-Gordan coefficients that appears in the scalar-multipole expressions (which we do using Mathematica) and some of the tensors $\bar{\mathcal Y}_L^{lm}$, which are given in Eq.~(2.12) of~\cite{Thorne:1980ru}.
To make the comparison for the quadrupole, we take each unique product of multipoles on the right-hand sides of Eq.~\eqref{eqn:flux-quad} and replace the scalar radiative mass and current multipoles with Eq.~\eqref{eqn: mode-multi mom relation }.
We then multiply by $a_l Y_{lm}$ and sum over $m$, so that we obtain the contribution of this term to $\mathcal U^{\mathcal F}_{ij} n^i n^j$.
We then compare this with the equivalent term (in terms of multipole order and number of time derivatives acting on the multipoles) in Eq.~\eqref{eqn:quadrupole PN exp decomposed} by contracting the expression there with $n^i n^j$, so as to verify that the two terms are equal.
By repeating this process for all the terms, we find that the expression for the quadrupole computed from Eq.~\eqref{eqn: flux contribution} [using the moments in Eq.~\eqref{eqn:flux-quad}] are equivalent to that in Eq.~\eqref{eqn:quadrupole PN exp decomposed}, up to the error terms involving $\ud^2/\ud u^2$ of products of canonical multipoles.
We repeat the same process for the octupole.

This then allows us to conclude that the following quadrupole and octupole fluxes can be written in the STF-tensor form, as we give below.
First, the octupole fluxes and pseudo-fluxes are
\begin{subequations} \label{eqn: octu leading term}
    \begin{align}
         (\mathcal{F}_{0})_{ijk} & = {} \biggr(\frac{G}{c^5}\biggr) \left(\frac{1}{2}\sqrt{\frac{5}{3}} M^{(3)}_{a\langle i} M^{(4)}_{jk\rangle a} -2\sqrt{\frac{3}{5}}\epsilon_{ab\langle i} M^{(3)}_{ja} S^{(3)}_{k\rangle b}\right), \label{eqn: F0 octu leading term} \\
        (\mathcal{F}_{1})_{ijk} & = -\frac{\sqrt{15}}{2} \biggr(\frac{G}{c^5}\biggr) \left(M^{(3)}_{a\langle i} M^{(3)}_{jk\rangle a}-M^{(2)}_{a\langle i} M^{(4)}_{jk\rangle a} \right), \label{eqn: F1 octu leading term} \\
       (\mathcal{G}_{2}^{\mathrm{rad}})_{ijk} & = - \biggr(\frac{G}{c^5}\biggr) \left(\frac{25}{2}\sqrt{\frac{5}{3}}M^{(2)}_{a\langle i}M^{(3)}_{jk\rangle a} +24\sqrt{15}\epsilon_{ab\langle i}M^{(2)}_{ja} S^{(2)}_{k\rangle b}\right), \label{eqn: G rad octu leating term} \\
       (\mathcal{G}_{2}^{\mathrm{nonrad}})_{ijk} & =  {} \sqrt{15} \biggr(\frac{G}{c^3}\biggr)  M M^{(3)}_{ijk} -\sqrt{15} \biggr(\frac{G}{c^5}\biggr) \left(5M^{(2)}_{a\langle i}M^{(3)}_{jk\rangle a} + 12 \epsilon_{ab\langle i}M^{(2)}_{ja} S^{(2)}_{k\rangle b} \right) . \label{eqn: G nonrad octu leating term}
    \end{align}
\end{subequations}
The relationship between the STF multipoles
$(\mathcal{F}_{n})_{L}$ and $(\mathcal{G}_{n})_{L}$ of the fluxes and pseudo-fluxes and the scalar multipoles $(\mathcal{F}_{n})_{lm}$ and $(\mathcal{G}_{n})_{lm}$ is the same as that given for the radiative mass multipole in Eq.~\eqref{eqn:radiative moments}:
\begin{equation}
    (\mathcal{F}_{n})_{L} = a_l \sum_{m=-l}^{l} (\mathcal{F}_n)_{lm}\mathcal{Y}^{lm} _{L}, \qquad
    (\mathcal{G}_{n})_{L} = a_l \sum_{m=-l}^{l} (\mathcal{G}_n)_{lm}\mathcal{Y}^{lm} _{L}.
\end{equation}
This implies that Eq.~\eqref{eqn: flux contribution} for the contribution of the fluxes and pseudo-fluxes to the mass octupole at 2.5~PN order relative to the leading octupole is given by
\begin{align} \label{eqn: flux octupole contribution}
    \mathcal{U}^{\mathcal{F}}_{ijk} = & -\frac{2}{\sqrt{15}}\int_{-\infty}^u \ud u_1\; (\mathcal{F}_{0})_{ijk}(u_1)+ \frac{1}{6\sqrt{15}}(\mathcal{F}_{1})_{ijk}(u) \nonumber\\
    &-\frac{1}{10\sqrt{15}} \frac{\ud}{\ud u}\left[ (\mathcal{F}_2)_{ijk} (u)+ (\mathcal{G}_{2}^{\mathrm{rad}})_{ijk}(u) + (\mathcal{G}_{2}^{\mathrm{nonrad}})_{ijk}(u)\right] + \frac{\ud^2}{\ud u^2} (\cdots).
\end{align}
Substituting the expressions for the fluxes from Eq.~\eqref{eqn: octu leading term} into Eq.~\eqref{eqn: flux octupole contribution} and comparing with Eq.~\eqref{eqn:octupole PN exp decomposed}, shows, in detail, how the PN expansion of the octupole moment can be understood in terms of the contributions from the flux and pseudo-flux pieces of the moments of the news. 
The PN memory integral comes from the zeroth flux $\mathcal{F}_0$, the instantaneous terms with no retarded-time derivatives come from the first flux $\mathcal{F}_1$ and both radiative and non-radiative the pseudo-flux pieces of the second moment contribute to the first total derivative term. 
The higher derivative terms could, in principle, be obtained from calculating flux and pseudo-flux contributions from the third moment of the news, although this still neglects the charge contribution to the third moment. 

Finally, note that the leading-order pseudo-flux term [the first term in Eq.~\eqref{eqn: G nonrad octu leating term}] appears at the same PN order as the tail term (the 1.5~PN order terms in the expansion of the radiative moments in \ref{sec: Appendix 1}).
The PN tail integral, however, contains both a logarithm term and a part without a logarithm; this latter term could be written as the total derivative of the mass monopole multiplied by $l+1$ derivatives of the $l$th canonical multipole.
The non-radiative pseudo-flux has the same form as this part of the tail term without the logarithm.
Such logarithm terms do not manifestly appear in any of the flux or pseudo-flux expressions, which may account for why it does not arise from our Bondi-Sachs calculations.
Thus, the tail terms should be arising from the charge contributions to the waveform, which we have not computed.

We can now repeat the calculation for the quadrupole.
By following the same approach that we used with the octupole, we can first determine that the STF form of the zeroth and first fluxes are 
\begin{subequations} \label{eq:quad-fluxes-STF}
\begin{align} \label{eqn: F0 mass quad leading terms}
    (\mathcal{F}_{0})_{ij} = {} & \frac{\sqrt{3}}{7} \biggr(\frac{G}{c^5}\biggr) \mathrm{M}^{(3)}_{a\langle i}\mathrm{M}^{(3)}_{j\rangle a} + \frac 1{7\sqrt 3} \biggr(\frac{G}{c^7}\biggr) \left(-\frac{5}{72}M^{(3)}_{ab} M^{(5)}_{ijab} + \frac{16}{3}  S^{(3)}_{a{\langle i}} S^{(3)}_{{j\rangle}a} \right.\nonumber\\
    & \left.+ \frac{10}{9}\epsilon_{ac{\langle i}}M^{(4)}_{{j\rangle}bc} S^{(3)}_{ab} -\frac{5}{4}\epsilon_{ac{\langle i}}S^{(4)}_{{j\rangle}bc} M^{(3)}_{ab}\right), \\
    (\mathcal{F}_{1})_{ij} = {} & \frac{5}{\sqrt 3} \biggr(\frac{G}{c^7}\biggr)  \left[ \frac{1}{72}(M^{(3)}_{ab} M^{(4)}_{ijab}-M^{(2)}_{ab} M^{(5)}_{ijab})+ \frac{2}{21}\epsilon_{ac{\langle i}}(M^{(4)}_{{j\rangle}bc} S^{(2)}_{ab}- M^{(3)}_{{j\rangle}bc} S^{(3)}_{ab}) \right.\nonumber\\
    & \left. - \frac{3}{28}\epsilon_{ac{\langle i}}(S^{(4)}_{{j\rangle}bc} M^{(2)}_{ab}-S^{(3)}_{{j\rangle}bc} M^{(3)}_{ab})\right]. \label{eqn: F1 mass quad  leading terms}
\end{align}
The STF multipolar expression for the pseudo-fluxes' contributions to the quadrupole are given by
\begin{align}
    (\mathcal{G}_{2}^{\mathrm{rad}})_{ij} = & -\frac{6\sqrt{3}}{7} \biggr(\frac{G}{c^5}\biggr) \mathrm{M}^{(2)}_{a\langle i}\mathrm{M}^{(2)}_{j\rangle a} 
    + \frac{1}{7\sqrt 3} \biggr(\frac{G}{c^7}\biggr)  \left(\frac{5}{54}M^{(2)}_{ab} M^{(4)}_{ijab} - \frac{25}{54} M^{(3)}_{ab{\langle i}}M^{(3)}_{{j\rangle}ab} \right. \nonumber\\
    &\left. +\frac{25}{2}\epsilon_{ac{\langle i}}S^{(3)}_{{j\rangle}bc} M^{(2)}_{ab}   + \frac{20}{9}\epsilon_{ac{\langle i}}M^{(3)}_{{j\rangle}bc} S^{(2)}_{ab} - 96 S^{(2)}_{a{\langle i}} S^{(2)}_{{j\rangle}a} \right),  \label{eqn: G rad mass quad  leading terms} \\
    (\mathcal{G}_{2}^{\mathrm{nonrad}})_{ij} = & \sqrt 3 \biggr(\frac{G}{c^3}\biggr) M M^{(2)}_{ij} -\frac{12\sqrt{3}}{7} \biggr(\frac{G}{c^5}\biggr) \mathrm{M}^{(2)}_{a\langle i}\mathrm{M}^{(2)}_{j\rangle a} 
    +\frac{5}{7\sqrt{3}} \biggr(\frac{G}{c^7}\biggr) \left( \frac 16 M^{(2)}_{ab} M^{(4)}_{ijab} \right. \nonumber \\
    &  \left. -\frac{5}{9}M^{(3)}_{ab{\langle i}}M^{(3)}_{{j\rangle}ab} -\frac{4}{3}\epsilon_{ac{\langle i}}M^{(3)}_{{j\rangle}bc} S^{(2)}_{ab} -\frac{3}{2}\epsilon_{ac{\langle i}}S^{(3)}_{{j\rangle}bc} M^{(2)}_{ab} \right) . \label{eqn: G nonrad quad  leading terms}
    \end{align}
\end{subequations}
This implies the full flux contributions to the quadrupole can be written in STF form as
\begin{align} \label{eqn: flux quadrupole contribution}
    \mathcal{U}^{\mathcal{F}}_{ij} = & -\frac{2}{\sqrt{3}}\int_{-\infty}^u \ud \tau (\mathcal{F}_{0})_{ij}(\tau) + \frac{1}{3\sqrt{3}}(\mathcal{F}_{1})_{ij}
    - \frac{1}{2\sqrt{3}} \frac{\ud}{\ud u}\left[(\mathcal{F}_2)_{ij} +(\mathcal{G}_{2} ^{\mathrm{rad}})_{ij} + (\mathcal{G}_{2} ^{\mathrm{nonrad}})_{ij}\right]. 
\end{align}
Substituting the STF multipole expressions for the fluxes and pseudo-fluxes in Eq.~\eqref{eq:quad-fluxes-STF} into Eq.~\eqref{eqn: flux quadrupole contribution} reproduces the waveform in Eq.~\eqref{eqn:quadrupole PN exp decomposed} up to the second-derivative terms. 
As with the octupole, the PN memory integral arises from the zeroth flux, the instantaneous terms without derivatives come from the first flux, and the second fluxes and pseudo-fluxes contribute to the term involving a single time derivative.
For the quadrupole case, the higher derivative terms cannot be obtained from flux and pseudo-flux contributions to higher moments of the news, because the angular operators which appear on left-hand side of Eq.~\eqref{eqn:shear_contribution elec} remove the $l = 2$ component of the shear for $n \geq 3$.
This follows from Eq.~\eqref{eqn:annihilator}: the derivative operator $\mathcal D_n$ acting on $Q_{n - 1}$ in the evolution equation for $Q_n$ sets the $l = n - 1$ mode to zero. 
This is why we do not include a residual term in Eq.~\eqref{eqn: flux quadrupole contribution}.
There also will be charge contributions from the $n = 2$ charge in the PN expression that we have been neglecting.

\section{Leading post-Newtonian contribution from the higher-memory moments for compact binaries} \label{sec: PN order of Moments}

We now specialize to non-spinning, compact-binary sources in quasi-circular orbits, and we write our results in terms of the post-Newtonian parameter $x$.
At leading Newtonian order, the PN parameter is given by $x=M/R$, with $R$ being the position of the reduced mass of the binary 
$M=m_1+m_2$ being the total mass, and $m_1$ and $m_2$ being the individual masses.
It will also be convenient to define $\eta = m_1 m_2/M^2$ as the symmetric mass ratio and $\delta m = m_1 - m_2$ to be the mass difference with the convention that $m_1 \geq m_2$.
We derive the leading PN-order contributions of the zeroth, first and second flux and pseudo-flux contributions to the moments of the news and the corresponding shear.
Some aspects of these calculations have been performed for the zeroth flux~\cite{Wiseman:1991ss} and first fluxes~\cite{Nichols:2017rqr,Nichols:2018qac} (in the sense that we will describe shortly below); however, they have not been computed for the second fluxes or pseudo-fluxes (though they are computed numerically in~\cite{Grant:2022bla} using full inspiral-merger-ringdown waveforms).

\subsection{Terminology for higher memory effects and calculation methods}

To describe the sense in which the flux calculations are not fully complete, we introduce a terminology of ``oscillatory'' and ``non-oscillatory'' higher memory effects.
The displacement memory effect for non-precessing binaries has the property that the leading $m=0$ modes generated by the zeroth flux are monotonically increasing.
These modes are described as ``non-oscillatory'' terms in the waveform, so as to distinguish them from the dominant oscillatory waveform $m\neq 0$ modes which have zero mean, when averaged over timescales short compared to the PN radiation-reaction timescale.
However, precisely the same zeroth flux terms that produce the $m=0$ modes in waveform also produce $m\neq 0$ modes that have zero mean.
This nonlinear contribution to the oscillatory waveform modes are what we refer to as the ``oscillatory'' memory terms.
Although in the PN context, they do not lead to an offset in the strain that accumulates from early times to late times, the oscillatory strain associated with these modes is computed from the same terms that generate the $m=0$ modes with an offset.
Thus, we still consider them to be ``memory'' modes, because they are both features of nonlinearities in the asymptotic Einstein equations.\footnote{These oscillatory modes can develop a nonzero offset during the merger and ringdown stages of a black-hole binary merger, even though during the early inspiral they have zero-mean; thus, they can have a memory offset in the standard sense, but the PN inspiral is not sufficient to calculate it (i.e., it must be computed from the full inspiral-merger-ringdown waveforms).}

In terms of the oscillatory and non-oscillatory memory effects, the higher memory effects associated with the first flux have been computed in the PN context for both types for the electric part of the first flux.
This is precisely the CM memory effect, which was computed at leading PN order in~\cite{Nichols:2018qac}.
However, only the leading PN non-oscillatory part of the flux has been computed for the magnetic part of the first flux in~\cite{Nichols:2017rqr}; the oscillatory terms have not been computed to the best of our knowledge.

We now summarize the computations required to compute the PN expressions for the moments of the news and the corresponding shear.
We use the PN expressions for the radiative modes given in Sec.~9.5 of~\cite{Blanchet:2013haa}.
We reproduce the relevant modes at the orders at which they are required in~\ref{app:radModesBinaries}, for $m\geq 0$ [the negative $m$ values can be obtained using the relationship that $h_{l(-m)} = (-1)^m\bar h_{lm}$].
These modes are written in terms of the mass parameters $M$, $\eta$ and $\delta m$; the PN parameter $x$; and the gravitational-wave phase $m \phi(x)$ for a given spherical-harmonic mode indexed by $(l,m)$.
To compute the electric-parity part of the quadrupole or octupole fluxes and pseudo-fluxes, we can directly substitute the relevant expressions for the modes in~\ref{app:radModesBinaries} into the expressions~\eqref{eqn:flux octu expansion} and~\eqref{eqn:flux-quad}.
For all the magnetic fluxes and pseudo-fluxes, as well as the $l\geq 4$ modes, we must calculate expressions analogous to~\eqref{eqn:flux octu expansion} and~\eqref{eqn:flux-quad} that give the spherical-harmonic modes of the scalar fluxes in terms of the radiative modes.
The calculations proceed analogously to those in Sec.~\ref{subsec:PNmomentsShear}, and the results are also lengthy expressions, such as those found in Eqs.~\eqref{eqn:flux octu expansion} and~\eqref{eqn:flux-quad}.
We, thus, do not give the expressions for general PN radiative modes; rather, we give only the leading-PN results for compact binaries below.

To extract the moment of the news, or the corresponding shear, we will compute integrals or derivatives of the fluxes and pseudo-fluxes.
Computing derivatives is straightforward: a derivative of the PN parameter $\ud x/\ud t \equiv \dot x$ is given by $\dot x = 64 \eta x^5/(5 M) + O(x^6)$; a derivative of the phase $\dot \phi$ only increases the PN order by $3/2$ [i.e., $\dot \phi = x^{3/2}/M + O(x^{5/2})$, which is just the Keplerian orbital frequency, at leading order].
Computing integrals to obtain the moments is somewhat more subtle.
As described in~\cite{Favata:2008yd} (and also~\cite{Nichols:2017rqr,Nichols:2018qac}), the integrals of the fluxes in time $t$ can be performed by recasting them in terms of $x$ (which introduces a factor of $1/\dot x \sim x^{-5}$).
The integrals of non-oscillatory terms have a PN order that is four PN-orders lower than the integrand, whereas the oscillatory terms have an order that is only three-halves PN-orders lower.
This counting allows the PN orders of moments to be estimated without too much difficulty.
However, it is also important to compute all of the relevant Clebsch-Gordan-type coefficients to determine whether the leading-PN coefficient of the flux is nonzero. 

We next summarize how we will present the results of our calculations.
The signals associated with the higher memory effects can all be written schematically as 
\begin{equation} \label{eq:hlm-hme-schematic}
    r \sqrt 2 h_{lm}^\mathrm{(HME)} = \hat h_{lm} x^{k+1} e^{-i m\phi} ,
\end{equation}
where $\hat h_{lm}$ is a coefficient proportional to the radiative modes $U_{lm}$ or $V_{lm}$ associated with the higher memory effects, with the phase $e^{-i m\phi}$ and the leading power of $x$ factored out.
Because we compute the leading-order effect for the zeroth, first and second fluxes and pseudo-fluxes, for both electric and magnetic parities, for both oscillatory and non-oscillatory effects, and for both the moment of the news and the associated shear, there would be nearly 40 different expressions of the form in Eq.~\eqref{eq:hlm-hme-schematic} to present.
We do not write these expressions explicitly.
Instead, we summarize the expressions in three tables.
Table~\ref{tab: Non Osci PN} contains the leading $l$ and $m=0$ non-oscillatory mode for each flux type, the corresponding PN orders [the power of $k$ in Eq.~\eqref{eq:hlm-hme-schematic}], and the oscillatory modes that contribute to the higher memory effects; Table~\ref{tab: Osci PN} contains the same information for the oscillatory modes.
Finally, Table~\ref{tab: Coefficients of Shear contribution } focuses just on the shear associated with the higher memory effects, and has the coefficient $\hat h_{lm}$ for the leading flux or pseudo-flux.
For the modes in Table~\ref{tab: Coefficients of Shear contribution }, it is then straightforward to reconstruct the corresponding radiative moments $U_{lm}$ or $V_{lm}$ [or $h_{lm}$ via Eq.~\eqref{eq:hlm-Ulm-Vlm}] from the results in Tables~\ref{tab: Non Osci PN}--\ref{tab: Coefficients of Shear contribution } and Eq.~\eqref{eq:hlm-hme-schematic}.
Similarly, the polarizations $h_+$ and $h_\times$ could be computed using the expression for $h_{lm}$ and Eq.~\eqref{eq:h-hplus-hcross} using the conventions for the spin-weighted spherical harmonics defined in~\ref{app:STFtensorIdentities}.

We also do not plot the higher-memory signals as a function of time or of the PN parameter $x$.
At leading Newtonian order, the time dependence of $x(t)$ and of the phase $\phi(x)$ are given by
\begin{equation}
    x(t) = \frac 14 \left[ \frac\eta{5M} (t_c-t)\right]^{-1/4} , \qquad
    \phi(x) = -\frac{x^{-5/2}}{32 \eta} .
\end{equation}
Thus, the flux or pseudo-flux contributions will look no different from a higher PN correction to a given $(l,m)$ spherical harmonic mode of the waveform.
Namely, the oscillatory signals will be chirp signals with a slowly increasing amplitude and frequency; the non-oscillatory signals will be just slowly increasing amplitudes.
The more complex features of these memory signals will arise during the last stages of the inspiral, the merger and the ringdown; however, the PN approximation does not hold in this regime, so we are unable to plot the waveform from the later stages of the merger using the PN results in this paper. Instead, see~\cite{Grant:2023ged}, particularly Figure~3, for explicit plots of some of these contributions in a numerical relativity simulation, which covers the late inspiral, merger and ringdown.

\subsection{Post-Newtonian orders and coefficients of higher memory effects}

The first two tables for the non-oscillatory (Table~\ref{tab: Non Osci PN}) and oscillatory (Table~\ref{tab: Osci PN}) higher memories give the flux or pseudo-flux (first column), the spherical-harmonic modes in which the higher memory appears (second column), the modes that are used in computing the higher memory (third column), the PN order of the moment of the news relative to the leading zeroth moment (fourth column), and the PN order of the contribution to the shear relative to the leading quadrupole waveform (fifth column).
Because both the zeroth moment and the shear scale as $x$, we wrote the mode schematically as $c_k x^{k+1} e^{-i m\phi}$, so that the PN order given in the last two columns of corresponds to the power $k$.
In Table~\ref{tab: Coefficients of Shear contribution }, we give the coefficient $c_k$ of the contribution of a moment or the news to the shear for the lowest $l$ and largest $m$ mode of a given parity (electric or magnetic).
We give the result for both the oscillatory and non-oscillatory modes, and we focus on the flux (or for the second moment, pseudo-flux) that has the lowest-PN-order contribution for a given moment of the news.
The reason for focusing on these specific contributions is related to the detection prospects for these effects, as we discuss in more detail later in this subsection.

\begin{table}[htb]
    \centering
    \caption{PN orders of the non-oscillatory modes relative to the leading $l=2$, $m=0$ mode.  
    Note that in the second column, the modes are electric ($U_{l0}$) or magnetic ($V_{l0}$) depending on whether $l$ is even or odd, respectively.
    This occurs because the binaries we consider have no spins and the orbital plane does not precess.
    The modes used to compute the flux or pseudo-flux are listed in the third column.
    The fourth and fifth columns gives the PN order (power $k$ of the term $x^{k+1}$) that arises in the expression for the moment of the news or the shear.} 
    \begin{tabular}{c l l l l} \hline\hline\hline
        Flux or & Leading-order modes & Contributing & \multicolumn{2}{c}{PN order of} \\\cline{4-5}
        pseudo-flux & in shear ($m = 0$) & modes & moment & shear \\\hline\hline
        $\mathcal{F}_0$ & $l = 2, 4$ & $U_{2(\pm 2)} $& $0$ & $0$ \\\hline\hline
        $\mathcal{F}_1$ & $l = 2, 4$ & \begin{tabular}{@{}l@{}} $U_{2(\pm 2)}, U_{4(\pm 2)}, U_{3(\pm 1)}$, \\ $V_{2(\pm 1)}, V_{3(\pm 2)}$ \\ \end{tabular} & ${1}$ & ${5}$ \\\hline
        $\mathcal{F}^*_1$ & $l = 3$ & $U_{2(\pm 2)} $ & ${-1.5}$ & ${2.5}$ \\\hline\hline
        $\mathcal{F}_2$ & $l = 2, 4, 6, 8$ & $U_{2(\pm 2)}, U_{20}, U_{40}$ & $2$ & ${10}$  \\\hline
        $\mathcal{F}^*_2$ & $l = 3, 5, 7$ & $U_{2(\pm 2)}, U_{20}, U_{40}$ & ${2}$& ${10}$ \\\hline
        $\mathcal{G}_2^\mathrm{rad}$ & $l = 2, 4$ &$U_{2(\pm2)}, U_{20}, U_{40}$ & ${-3}$ & $5$ \\\hline
        $\mathcal{G}^{* \mathrm{rad}}_2$ &$l=3,5$  &\begin{tabular}{@{}l@{}} $U_{2(\pm2)},U_{3(\pm1)},U_{4(\pm2)}$, \\ $V_{2(\pm1)}, V_{3(\pm2)}$ \\ \end{tabular} & $-0.5$  & $7.5$  \\\hline
        $\mathcal{G}_2^\mathrm{nonrad}$ & $l = 2, 4$ &$U_{20}, U_{40}$ & ${-4}$ & $4$ \\\hline
        $\mathcal{G}^{* \mathrm{nonrad}}_2$ & $l = 3$ & $V_{30}$ & ${-1.5}$ & ${6.5}$ \\\hline\hline\hline
    \end{tabular}
    \label{tab: Non Osci PN}
\end{table}

In Table~\ref{tab: Non Osci PN}, there are a few features that we comment upon. 
First, the expressions for the zeroth flux and the magnetic part of the first flux are consistent with previous calculations for the displacement memory (e.g.,~\cite{Wiseman:1991ss,Favata:2008yd,Nichols:2017rqr}) and the spin memory effects~\cite{Nichols:2017rqr}.
The expressions the non-oscillatory electric first flux (second row) and for the second fluxes and pseudo-fluxes in the last six rows are new.
The order of the zeroth moment and shear are the same, because the zeroth moment of the news is precisely the change in the shear.
The PN orders of the first fluxes' moments and shear differ by four PN orders, because differentiating the moment once replaces one factor of $x$ with $\dot x \sim x^5$, which leads to a net increase of four powers of $x$.
This also explains why the moments associated with the second fluxes and pseudo-fluxes are eight PN orders higher.
The shear associated with the second moments involves taking two derivatives of the moment; the first derivative increases the power of $x$ by four, and the second derivative increases it by four again.
For the non-oscillatory shear associated with the second fluxes and pseudo-fluxes, the non-radiative pseudo-flux $\mathcal G_2^\mathrm{nonrad}$ has the lowest PN contribution (at 4~PN).
Although the 4~PN waveform has been computed for the $l=2$, $m=\pm 2$ waveform, the contribution to the $l=2$, $m=0$ mode has only been computed at 1.5~PN order (in terms of the parameter $x$) for compact binaries in~\cite{Blanchet:2023sbv}.
This prevents us from making a comparison with existing computations in this case.
As we also discuss in more detail below, this non-radiative pseudo-flux term is also closely related to the charge contribution to the CM memory effect, which was computed in~\cite{Nichols:2018qac}.

The PN order of the contributions of these fluxes to the shear is an important contributing factor to determining if the higher memory effects will be detectable. 
The zeroth moment or the displacement memory, which appears in the waveform at leading Newtonian order, can potentially be observed in current ground-based detectors (in the entire population of black-hole mergers~\cite{Lasky:2016knh,Boersma:2020gxx,Grant:2022bla}) or in next-generation ground-based detectors~\cite{Grant:2022bla} for individual events. 
The magnetic part of the shear generated by the first moment (the spin memory) enters at 2.5~PN order: it was recently forecast that it may be detectable by next-generation detectors only in the entire population of black hole mergers after several years of observation time at the detectors' design sensitivities~\cite{Grant:2022bla}. 
The electric part, the CM memory, is a relative 5~PN order effect (for the non-oscillatory part).
Given that it is another 2.5~PN orders higher than the spin memory effect, this does not bode well for its detection prospects; however, tidal effects in neutron-star binaries behave like a 5~PN correction to the gravitational-wave phase, and they have been constrained by individual gravitational-wave measurements~\cite{LIGOScientific:2017vwq}, because the coefficient of the effective 5~PN term is large.
Thus, computing the coefficient is another important factor in determining the detection prospects.
We will discuss this in more detail in the context of Table~\ref{tab: Coefficients of Shear contribution }.
For the second moment, or ballistic memory, the flux contribution is higher order (10~PN for both electric and magnetic parts); however, the contributions from the pseudo-fluxes are lower (4 and 6.5~PN for the electric and magnetic pieces respectively).
The high PN-order of these effects again makes their detection prospects seem more pessimistic, but we will comment on this further in the context of Table~\ref{tab: Coefficients of Shear contribution }.

\begin{table}[htb]
    \centering
    \caption{PN order of the oscillatory modes relative to the leading $l=2$, $m=\pm 2$ modes.
    The table is otherwise the same as Table~\ref{tab: Non Osci PN}, except that we also list the azimuthal spherical-harmonic index $m$ in parentheses in the second column and the modes in this column are $U_{lm}$ when $l+m$ is even and $V_{lm}$ when $l+m$ is odd.}
    \begin{tabular}{c l l l l} \hline\hline\hline
        Flux or & Leading-order & Contributing & \multicolumn{2}{c}{PN order of} \\\cline{4-5}
        pseudo-flux & modes in shear & modes & moment & shear \\\hline\hline
        $\mathcal{F}_0$ & $l = 4$ ($m = \pm 4$) & $U_{2(\pm 2)}$ & ${2.5}$ & ${2.5}$ \\\hline\hline
        $\mathcal{F}_1$ & \begin{tabular}{@{}l@{}} $l = 3$ ($m = \pm 1, \pm 3$), \\ $l = 5$ ($m = \pm 1, \pm 3, \pm 5$), \\ $l = 7$ ($m= \pm 1, \pm 3$) \end{tabular} & \begin{tabular}{@{}l@{}} $U_{3(\pm 1)}, U_{3(\pm 3)}, U_{2(\pm 2)}$, \\ $U_{20}, U_{40}$ \end{tabular} & ${1.5}$& ${3}$ \\\hline
        $\mathcal{F}^*_1$ & $l = 3$ ($m = \pm 2$) & $U_{2(\pm 2)}$, $U_{20}, U_{40}$ & ${1}$ & ${2.5}$ \\\hline\hline
        $\mathcal{F}_2$ & \begin{tabular}{@{}l@{}} $l = 2$ ($m = \pm 2$), \\ $l = 4$ ($m = \pm 2, \pm 4$), \\ $l = 6$ ($m = \pm 2, \pm 4, \pm 6$), \\ $l = 8$ ($m = \pm 4$) \end{tabular} & $U_{2(\pm 2)}, U_{20}, U_{40}$ & $2$ & ${5}$  \\\hline
        $\mathcal{F}^* _2$ & \begin{tabular}{@{}l@{}} $l = 3$ ($m = \pm 2$), \\ $l = 5$ ($m = \pm 2, \pm 4$), \\ $l = 7$ ($m = \pm 4$) \end{tabular} & $U_{2(\pm 2)}, U_{20}, U_{40}$& ${2}$ & ${5}$  \\\hline
        $\mathcal{G}^\mathrm{rad}_2$ &\begin{tabular}{@{}l@{}}$l=2$ ($m=\pm2$),\\ $l = 4$ ($m = \pm 4,\pm 2$) \end{tabular} & $U_{2(\pm 2)}, U_{20}, U_{40}$ & ${-.5}$ & ${2.5}$ \\\hline
        $\mathcal{G}^{*\mathrm{rad}}_2$ &\begin{tabular}{@{}l@{}} $l = 2$ ($m = \pm 1$),\\ $l = 4$ ($m = \pm 1, \pm 3$) \end{tabular} & \begin{tabular}{@{}l@{}} $U_{2(\pm 2)}, U_{3(\pm1)}, U_{3(\pm3)}$, \\ $V_{2(\pm1)}$ \end{tabular} & ${0}$ & ${3}$ \\\hline
        $\mathcal{G}^\mathrm{nonrad}_2$ & $l = 2$ ($m = \pm 2$) & $U_{2(\pm 2)}$ & ${-1.5}$ & ${1.5}$ \\\hline
        $\mathcal{G}^{*\mathrm{nonrad}}_2$ & $l = 2$ ($m = \pm 1$) & $V_{2(\pm 1)}$& ${-1}$ & ${2}$ \\\hline\hline\hline
    \end{tabular}
    \label{tab: Osci PN}
\end{table}

We now turn to discussing the oscillatory (higher) memory terms summarized in Table~\ref{tab: Osci PN}.
The oscillatory displacement memory (zeroth flux) is both at a higher PN order than the non-oscillatory part and it appears in the higher $l=4$ multipole moment.
For the first fluxes, the corresponding moments and the shear differ by only 1.5 PN orders (as described earlier).
For the electric part, this makes the oscillatory shear a lower PN-order expression than the corresponding term for the non-oscillatory modes, whereas the converse is true for the moment.
The $U_{40}$ mode contributing to the flux was neglected in the calculation in~\cite{Nichols:2018qac}, which is why no $l=7$ modes were computed there; neglecting this term also changes the amplitude of the $l=3$ and $l=5$ terms.
For the magnetic part, the contribution to the shear for the oscillatory and non-oscillatory terms enters at the same PN order (which implies that the non-oscillatory term appears at a lower PN order in the moment).
For the second fluxes and pseudo-fluxes, the shear is only three PN orders higher than the corresponding moments.
Comparing the oscillatory and non-oscillatory terms for the moments, the non-oscillatory terms are mostly at equal or lower PN orders than the corresponding oscillatory terms; however, for the shear, the converse is true for all of the terms: the oscillatory terms are all at lower PN orders than the equivalent non-oscillatory terms.
In fact, the non-radiative pseudo-flux terms arise in the shear at a lower PN order than the first flux terms.

The detection prospects of these oscillatory higher memory effects merits further discussion.
It was noted in~\cite{Nichols:2018qac} that the oscillatory part of the first flux could, in principle, be detectable, but it would just be a part of the total oscillatory spherical-harmonic mode (which in the language used in this paper, has substantial charge contributions).
At first glance, the same could be true of the non-radiative pseudo-flux term appearing at 1.5~PN order in the $l=2$, $m=\pm 2$ modes of the waveform.
This term also seems promising from the perspective of observation, because it is at a lower PN order, and it appears in the dominant quadrupole mode for compact binaries.
To have a more quantitative viewpoint on its detection prospects, however, it is important to know the size of the coefficient that multiplies this 1.5~PN contribution to the shear.
We discuss this next.

In Table~\ref{tab: Coefficients of Shear contribution }, we show the coefficient of the leading-order contribution to the non-oscillatory and oscillatory pieces of the shear's radiative moments in the second and third principal columns of the table.
We include only the lowest $l$ and highest $m$ term when there are multiple $(l,m)$ modes at a given PN order.
We include the zeroth, first and second flux and pseudo-flux of the moments.
For reference, the coefficient of the leading $(2,2)$ mode is $8\sqrt{2\pi/5} M\eta$.

\begin{table}[htb]
    \centering
    \caption{Coefficient of the leading-order flux and pseudo-flux contributions to the radiative multipoles of the shear.
    As in Table~\ref{tab: Osci PN}, the modes are $U_{lm}$ when $l+m$ is even and $V_{lm}$ when $l+m$ is odd.
    For brevity, we have only included the highest $m$ for the lowest $l$ terms, with $m > 0$ for the oscillatory contributions. 
    We do not include the PN order (given in Tables~\ref{tab: Non Osci PN} and~\ref{tab: Osci PN}) or the overall phase dependence $e^{-im\phi}$ (which can be inferred from the azimuthal number $m$ of the mode).}
    \begin{tabular}{c c c c c c} \hline\hline\hline
        Flux or &\multicolumn{2}{c}{Non-oscillatory piece} & & \multicolumn{2}{c}{Oscillatory piece} \\\cline{2-3}\cline{5-6}
        pseudo-flux & Mode & Coefficient & & Mode & Coefficient \\\hline\hline
        ${\mathcal{F}_0}$ & $(2,0)$ & {$\frac{4}{7}\sqrt{\frac{5\pi}{3}}M\eta$} & \vphantom{$\displaystyle\int$} & $(4,4)$ &$-\frac{8 i}{45}\sqrt{\frac{2\pi}{7}} M \eta^2$  \\\hline\hline % e^{-4i\phi}
        ${\mathcal{F}_1}$ & $(2,0)$ & \tablefootnote{The expression for this coefficient is somewhat lengthier than the others. It is given by \begin{equation}\label{eqn: COM U20} \frac{2}{315M}\sqrt{\frac{\pi}{15}}\eta^2\biggr[80M^2(-121+538\eta)-3\delta m^2(23+20 \ln 2) \biggr].\end{equation} The reason for the longer expression is that the leading-order terms in the modes given in Table~\ref{tab: Non Osci PN} vanish, but subleading terms do not. These subleading terms all appear at 2.5~PN order relative to the quadrupole moment, and they are out of phase by $\pi/2$ with the leading modes (because they arise from dissipative, radiation-reaction effects). The product of these subleading terms with the leading moments does not cancel.} & \vphantom{$\displaystyle\int$} & $(3,3)$ & {$-\frac{509}{990}\sqrt{\frac{\pi}{21}} \delta m\eta^2 $} \\\hline %e^{-i\phi}
        ${\mathcal{F}^*_1}$ & $(3,0)$ &{$-\frac{96}{5}\sqrt{\frac{\pi}{105}} M\eta^2$} & \vphantom{$\displaystyle\int$} & $(3,2)$&{$\frac{233}{405}\sqrt{\frac{2\pi}{7}} M\eta^2 $} \\\hline\hline %e^{-2i\phi}
        ${\mathcal{G}^{\mathrm{rad}}_2}$ & $(2,0)$ & $-\frac{16084864}{46305} \sqrt{\frac{\pi}{15}}M \eta^3$&\vphantom{$\displaystyle\int$}& $(2,2)$ & $-\frac{926i}{567}\sqrt{\frac{2\pi}{5}} M \eta^2$ \\ \hline
        ${\mathcal{G}^{\mathrm{rad}\,^*}_2}$ & $(3,0)$ &\tablefootnote{This expression is also longer, for the same reasons discussed in the previous footnote. It is given by \begin{equation}\label{eqn: Grad* U30}
            \frac{512}{25} \sqrt{\frac{\pi }{105}} (424 \eta -97) {M\eta}^3.
        \end{equation} } &\vphantom{$\displaystyle\int$} & $(2,1)$ &$-\frac{8749 i}{1701}\sqrt{\frac{\pi}{10}}\delta m \eta^2$\\ \hline
        ${\mathcal{G}^{\mathrm{nonrad}}_2}$ & $(2,0)$ &{$-\frac{128}{7}\sqrt{\frac{\pi}{15}}M\eta^2  $} & \vphantom{$\displaystyle\int$} & $(2,2)$ &{$-4i \sqrt{10 \pi}M\eta  $}\\\hline %e^{-2i\phi}
        ${\mathcal{G}^{\mathrm{nonrad}\,^*}_2}$ & $(3,0)$ &{$-\frac{512}{125}\sqrt{\frac{21\pi}{5}}M\eta^3 $} & \vphantom{$\displaystyle\int$} & $(2,1)$ &{$\frac{4i}{3}\sqrt{\frac{2\pi}{5}}\delta m \eta  $} \\\hline\hline\hline %e^{-i\phi}
    \end{tabular}
    \label{tab: Coefficients of Shear contribution }
\end{table}

The non-oscillatory zeroth flux coefficient is consistent with the previous results computed for the displacement memory, as is the non-oscillatory first magnetic flux terms.
The oscillatory, electric-parity, first-flux coefficient corrects the result in~\cite{Nichols:2018qac}, because it includes the $U_{40}$ mode which was dropped in~\cite{Nichols:2018qac}.
The oscillatory zeroth flux, non-oscillatory first electric flux and oscillatory magnetic first flux terms have not appeared elsewhere, as far as we are aware.
All of the second pseudo-flux terms have not been computed previously through this prescription.
However, the $(2,0)$ mode corresponding to the non-radiative electric pseudo-flux contribution has exactly the same form as the ordinary part of (i.e., charge contribution to) the CM memory effect computed in~\cite{Nichols:2018qac}, in the context of stationary-to-stationary transitions in asymptotically flat spacetimes.
This occurs because, in the context of stationary-to-stationary transitions, the value of the angular momentum aspect is computed from the expression for the non-radiative pseudo-flux in a stationary region with nonvanishing shear.
Nevertheless, the procedure used here does not make the stationary-to-stationary assumption, and still arrives at the same result.
However, there is an additional subtlety that we discuss next in the context of the oscillatory non-radiative electric pseudo-flux term in the $(2,2)$ mode.

This oscillatory term appears at 1.5~PN order relative to the leading quadrupole waves, which appears at precisely the same order as the tail terms.
However, the coefficient of the tail contribution is precisely $2\pi$ times larger than the leading $U_{22}$ mode for nonspinning compact binaries in quasi-circular orbits, whereas this pseudo-flux term is $-5i/2$ times larger than the leading $U_{22}$.
Notably, the factor of $-i$ implies that it is $-\pi/2$ out of phase with the tail term.
The fact that the 1.5~PN order waveform has a strictly real coefficient strongly suggests that there are contributions from the charge that cancel with this term.
In fact, from the first term of Eq.~(2.53d) of~\cite{Blanchet:2023pce}, one can see (after accounting for the differences in notation) that there is a contribution to $\ddot Q_2$ that is precisely equal to this contribution from the non-radiative pseudo-flux term.
This suggests that one should be cautious about computing the flux terms in isolation, especially for the higher fluxes, because the charge terms can have comparable contributions.

Thus, while the electric non-radiative pseudo-flux terms appeared nominally to be a promising gravitational-wave signature associated with the second moment of the news, they ultimately do not lead to any measurable effect, because they are precisely balanced by the ordinary (charge) contributions to the strain.
This is true of both the oscillatory terms (at the same order as the gravitational-wave tails) and the non-oscillatory term (which was previously identified as the ordinary contribution to the CM memory in~\cite{Nichols:2018qac} for stationary-to-stationary transitions) for the electric part of the shear.
While Ref.~\cite{Blanchet:2023pce} has not computed the contributions from the current moments, we strongly suspect that the same will hold for the magnetic non-radiative fluxes, although such a result has not been computed explicitly yet.

\section{Conclusions and Discussion} \label{sec:conclusions}

\subsection{Summary and conclusions}

In this paper, we have studied generally, and in the context of PN theory and compact binaries, the first three orders in a hierarchy of memory effects, which were introduced in~\cite{Grant:2021hga}, in the setting of asymptotically flat spacetimes.
We used the phrase ``higher memory effects'' to describe the collection of these effects, which are examples of persistent gravitational-wave observables.
In asymptotically flat spacetimes, these higher memory effects are encoded in the temporal moments of the Bondi news.
Specifically, the zeroth and first moments contain equivalent information as the displacement, spin and CM memories; we also studied the second moment, which has recently been dubbed the ballistic memory~\cite{Grant:2023ged}.

One main result of this paper was expressing all of the scalar moments of the news and of its dual in terms of a convenient hierarchy of integrals of fluxes (and pseudo-fluxes) and changes in a charge for that moment.
We used the nomenclature that charges are superpositions of Bondi aspects, fluxes are nonlinear terms that vanish in the absence of radiation, and that other nonlinear terms that are non-zero in a non-radiative region, but involve the shear and/or the Bondi aspects, were called pseudo-fluxes. 
Because the higher moments are computed from additional integrals of fluxes (and pseudo-fluxes) that arise in the calculation of lower moments, we focused primarily on the ``new'' flux and pseudo-flux terms that first appear in a given expression for a moment of the news.
We also identified contributions to the shear that are generated by the fluxes and charges for each moment of the news; the shear is the observable that can be measured most straightforwardly in interferometric gravitational-wave detectors, which makes it the more natural quantity to consider in the context of gravitational-wave detection and data analysis.

A second key result was the computation of the multipolar expansion of all of the flux and pseudo-flux terms that appear in the expressions for the zeroth, first and second moments of the news, in terms of the multipolar expansion of the shear.
These results hold generally, but our main application of them in this paper was to compute post-Newtonian expressions for the flux contributions to the moments of the news and the corresponding shear, both for general multipolar, PN sources and for nonspinning, quasicircular compact binaries.

We used the multipolar expansion of the zeroth, first and second fluxes (and pseudo-fluxes) to compute the contributions to the gravitational waveform for the mass quadrupole and octupole moments through 3.5~PN order for the quadrupole and 3~PN order for the octupole.
These fluxes could account for both the terms called ``memory'' in the PN literature and the large number of instantaneous, nonlinear terms.
The discrepancies between the two expressions were restricted to the tail terms (which cannot obviously be captured by using leading-order PN waveforms in the fluxes) and second derivative terms that come from charge terms, as well as the third fluxes and pseudo-fluxes for the octupole.
Given the differences between the harmonic-gauge equations of motion and the Bondi ones, the high degree of similarity of the final expressions is notable.

We finally computed the leading radiative PN multipoles that are produced by both the electric and magnetic zeroth, first and second fluxes and pseudo-fluxes for non-spinning, quasi-circular compact binaries.
Our calculations using zeroth and first fluxes reproduced the previously computed results for the displacement and spin memory effects, respectively, and we also corrected a small error in a previous computation of the center-of-mass memory effect with the first-flux computation.
The second fluxes produced a shear, for the non-oscillatory $m=0$ modes, at 10~PN order relative to the leading quadrupole.
The oscillatory shear associated with the second flux appears at 5~PN order.
Although the order is the same as tidal effects, it is a correction to the amplitude, rather than the phase, which would make it more challenging to detect.
There were also intriguing 1.5~PN order terms that arose from the electric non-radiative pseudo-flux.
However, these terms would be canceled by equivalent terms that appear in the charge contribution to the shear, so they cannot account for the tail terms at 1.5~PN order that have previously been computed.

\subsection{Discussion and future directions}

Much of the emphasis of the calculations in this paper was on determining the lowest multipole moments (quadrupole and octupole) both to compare with existing PN calculations and to determine the size of the new flux contributions to the moments relative to the leading part of the gravitational-wave strain for compact binary sources.
These leading-order results could be generalized to include sub-leading corrections from higher PN terms in the oscillatory waveforms.
The assumption of non-spinning compact objects could also be revisited to determine additional subleading corrections that enter from including spins.
As one of the main goals was to compute the leading-order effects to simply determine which of the many flux terms merit further study, we will leave calculations with spinning PN results (which are more involved computationally) to future studies.

Understanding the largest higher-memory flux terms is a pre-requisite for assessing the detection prospects of these new contributions to the higher moments of the news.
Such forecasts for the lower moments in ground-based detectors have already appeared in the literature.
Because the flux contributions are small compared to the dominant quadrupole modes, this is typically done by ``stacking''~\cite{Lasky:2016knh}.
For studies using the binary-black-hole mergers discovered to date by LIGO and Virgo~\cite{Hubner:2019sly, Hubner:2021amk}, the stacking involves combining the evidence ratio of models that include these contributions versus those that do not.
For forecasts of measurement prospects, it is also useful to consider stacking the signal-to-noise ratio of the memory signals in quadrature~\cite{Lasky:2016knh,Boersma:2020gxx}, which approximates the total evidence ratio (for all merger events) with and without the higher memories (see~\cite{Grant:2022bla} for more details about when this equivalence holds).

The outcomes of these studies are that the flux contribution from the zeroth moment (the nonlinear displacement memory) has not yet been detected in ground-based detectors~\cite{Hubner:2019sly, Hubner:2021amk}, but the prospects for its detection are promising~\cite{Lasky:2016knh,Boersma:2020gxx,Grant:2022bla}; similarly, for the magnetic part of the first moment's flux contribution (i.e., the spin memory), it is possible for the next generation of ground-based detectors to observe the statistical evidence for the memory in the population of merging black holes~\cite{Grant:2022bla}.

As future work, it is natural to investigate if similar studies could be done for the new higher memory effects arising from the second fluxes or pseudo-fluxes. 
The results of this paper suggest that the oscillatory pieces of the non-radiative pseudo-flux would be the most promising contribution given its relatively low PN order (2.5~PN), and the fact that it appears in the dominant $l=2$, $m=2$ waveform mode in non-spinning compact binaries.
Based on the facts that it has the same PN order as the spin memory effect and that the amplitudes of the effects have the same mass dependence and numerical order of magnitude, this suggests that it might also be detectable in the population of black-hole mergers by the next generation of ground-based GW detectors.
For example, since the spin memory has a peak strain on the order of $10^{-24}$ for a GW150914-like event, and an SNR in Advanced LIGO of order 0.01~\cite{Nichols:2017rqr}, it is not unreasonable to suppose the same would be true for this new radiative pseudo-flux term.
To obtain the most accurate forecasts in this context, it would be better to use the full inspiral-merger-ringdown signals from numerical relativity simulations rather than the inspiral-only PN signals considered in this paper.
This is an important reason why we leave this effort for future work.

For tests that involve just the post-Newtonian signal, one could envision making use of the parametrized tests of general relativity performed by the LIGO-Virgo-KAGRA Collaboration  (see, e.g.,~\cite{LIGOScientific:2021sio}).
These tests introduce ``deviation parameters'' in the coefficients of the post-Newtonian expansion of the gravitational-wave phase in the frequency-domain that have the property that when they are zero, the waveform reduces to the expression in general relativity.
Current constraints on these deviation parameters are consistent with zero, to within the statistical uncertainty of the measurement of these parameters.
If future measurements constrain the parameters to have a statistical error less than the contribution of the radiative pseudo-flux to the phase, then this would provide indirect evidence that the pseudo-flux contribution to the waveform exists: namely, because the pseudo-flux term is a prediction of general relativity that contributes to the value of the PN coefficient in general relativity.

There are some subtleties that make this using this approach a bit more difficult than it might, at first consideration, seem.
When writing the full $l=2$, $m=2$ waveform mode in terms of its time-dependent slowly varying amplitude and rapidly varying phase, the pseudo-flux contribution to the waveform affects the amplitude at 2.5~PN order, and it is imaginary rather than real.
While changes in the amplitude can produce phase differences when using energy balance to take into account the effect of the radiative losses on the binary's dynamics, these imaginary terms in the amplitude first produce phase differences at twice the PN order (namely 5~PN) unlike the real terms in the amplitude that change the phasing at the same PN order that they appear in the amplitude.
Currently, however, the parametrized tests of general relativity are restricted to 3.5~PN order and lower terms in the phase; they would need to be extended to higher PN orders (once these orders have been computed) to be able to obtain such indirect evidence of the pseudo-flux terms.

Although ground-based gravitational-wave detectors have larger calibration uncertainties on amplitude than on phase (see, e.g.,~\cite{LIGOScientific:2017aaj}) and tests of general relativity tend to be more sensitive to deviations in phase than in amplitude at a given PN order (see, e.g.,~\cite{Tahura:2019dgr}), performing tests of general relativity with deviations in the amplitude is still possible. 
Given the fact that the contributions from the radiative pseudo-flux terms appear at 2.5~PN order in the amplitude and 5~PN order in the phase, the significantly lower PN order in the phase could make up for the decreased constraining power of amplitude (as opposed to phase) effects.
Even if a parametrized test of relativity using the amplitude were implemented, to find evidence for the pseudo-flux contribution to the 2.5~PN waveform amplitude, the test would need to be performed at a high precision, because the pseudo-flux contribution is small.
Specifically, the part of the amplitude that depends on $\eta^2$ has a real and an imaginary part with comparable magnitudes; the pseudo-flux contribution is less than one-percent of the imaginary part. 
It would likely require either a large number of events or very high signal-to-noise events (and perhaps both) to show that the 2.5~PN deviation parameter is consistent with general relativity to less than one-percent accuracy (and thus the pseudo-flux is required to reproduce the observed waveform).
Thus, the first method, using inspiral-merger-ringdown waveforms to make forecasts may be the more promising future direction.

From a more theoretical perspective, the calculations in this paper also highlighted which fluxes are necessary to compute a given multipolar order of the gravitational-wave strain.
In particular, the quadrupole can be computed from just the zeroth, first and second fluxes (and pseudo-fluxes), plus a charge contribution.
The octupole also requires the third fluxes and a charge contribution that involves more Bondi aspects than the quadrupole.
This pattern continues for the $l$-pole mode of the strain, which involves $l+1$ flux terms and a charge with $l+1$ Bondi aspects.
The higher multipoles thus contain a larger set of possible nonlinear terms which are related to higher-order (or more subleading in $1/r$) parts of the Bondi-gauge metric and Einstein equations (cf.\ the discussion of ``memoryful'' and ``memoryless'' charges in~\cite{Compere:2022zdz}).
This is not as apparent in the PN calculations of the waveform, in which, at a given PN order, the quadrupole often appears to be the most complicated mode.
This occurs because the higher multipoles are also of a higher PN order than the quadrupole as are the higher fluxes; thus the higher flux contributions are at higher PN orders than the current state-of-the-art PN calculations.
This perspective from the Bondi-Sachs framework also provides one more motivation for investigating higher gravitational-wave multipoles.

\section*{Acknowledgments}

D.A.N.\ and S.S.\ acknowledge support from NSF Grants No.\ PHY-2011784 and No.\ PHY-2309021.
A.M.G.\ acknowledges the support of the Royal Society under grant number {RF\textbackslash ERE\textbackslash 221005}.

%\clearpage
\section*{References}
\bibliographystyle{iopart-num}
\bibliography{refrences}

\appendix

\section{PN expansion of the mass quadrupole and octupole  moments} \label{sec: Appendix 1}

The flux contributions to the shear from the higher moments of the news enter as $n$ derivatives with respect to retarded time for the n\emph{th} flux [see Eq~\eqref{eqn: flux contribution}]. 
In addition, the fluxes in the Bondi approach involve only the radiative multipoles, which involve $l$ retarded-time derivatives of the multipole moment of order $l$; however, the post-Newtonian calculation involves the full metric which includes the multipole moments without any derivatives (which would correspond to as many as $l$ integrals of the radiative multipole of order $l$).
Thus, to compare the expression for the flux contribution to the Bondi radiative moments [Eqs.~\eqref{eqn: flux octupole contribution} and~\eqref{eqn: flux quadrupole contribution}] with the PN radiative moments we will need to re-express the PN results (given in Sec.~2.4.5 of~\cite{Blanchet:2013haa}, for example).
We use the product rule so as to write it in a form that appears more compatible with the Bondi framework.
The result of these calculations is that the PN waveform can be written in a form involving integrals over retarded time, ``instantaneous'' terms without derivatives, and terms that involve one or two total derivatives (for the quadrupole and octupole). 
In addition, we require that the terms that involve zero or one time derivative do not involve multipole moments that would correspond to retarded time integrals of the Bondi radiative moments.
This allows the results below in Eqs.~\eqref{eqn:quadrupole PN exp decomposed} and~\eqref{eqn:octupole PN exp decomposed} to be compared with the respective Bondi results in Eqs.~\eqref{eqn: flux quadrupole contribution} and~\eqref{eqn: flux octupole contribution}, respectively.
For the mass quadrupole moment, the result is
\begin{align} \label{eqn:quadrupole PN exp decomposed}
    \mathcal{U}_{ij} =& M^{(2)}_{ij} + 2M \left(\frac{G}{c^3}\right)\int_{-\infty} ^U \ud \tau \biggr[\ln\left(\frac{c(U-\tau)}{2r_0}\right) + \frac{11}{12}\biggr] M^{(4)}_{ij}(\tau)\nonumber\\ 
    &+\left(\frac{G}{ c^5}\right)\biggl[  -\frac{2}{7}\int^{U}_{-\infty} \ud \tau\,
    \mathrm{M}^{(3)}_{a\langle i}\,\mathrm{M}^{(3)}_{j\rangle a}(\tau)\nonumber\\  
    &+ \frac{9}{ 7}\frac{\ud}{\ud u}\left(M^{(2)}_{a\langle
    i}M^{(2)}_{j\rangle a}\right) + \frac{\ud^2}{\ud u^2}\left( \frac{1}{7} M^{(3)}_{a\langle i}M_{j\rangle a}- M^{(2)}_{a\langle
    i}M^{(1)}_{j\rangle a} 
    +\frac{1}{3}\epsilon_{ab\langle i}M^{(2)}_{j\rangle
    a}S_{b}\right)\biggr]\nonumber \\ 
    & + \left(\frac{G}{c^7}\right) \left\{ -\frac{32}{63}
    \int^{U}_{-\infty} \ud \tau\, \mathrm{S}^{(3)}_{a{\langle
        i}}\,\mathrm{S}^{(3)}_{{j \rangle}a}(\tau) + \frac{5}{756}
    \int^{U}_{-\infty} \ud \tau\,
    \mathrm{M}^{(3)}_{ab}\,\mathrm{M}^{(5)}_{ijab}(\tau) \right.
    \nonumber\\ 
    & \qquad\quad - \frac{20}{189}
    \,\epsilon_{ab{\langle i}} \int^{U}_{-\infty} \ud \tau\,
    \mathrm{S}^{(3)}_{ac}\,\mathrm{M}^{(4)}_{{j \rangle}bc}(\tau) +
    \frac{5}{42} \,\epsilon_{ab{\langle i}} \int^{U}_{-\infty}
    \ud \tau \,\mathrm{M}^{(3)}_{ac}\,\mathrm{S}^{(4)}_{{j \rangle}bc}(\tau) \nonumber \\ & \qquad\quad+ \frac{5}{648}(M^{(3)}_{ab} M^{(4)}_{ijab} -M^{(2)}_{ab} M^{(5)}_{ijab} )+\epsilon_{ac{\langle i}} \biggr[-\frac{5}{84}(S^{(4)}_{{j\rangle}bc} M^{(2)}_{ab}-S^{(3)}_{{j\rangle}bc} M^{(3)}_{ab})\nonumber \\
    & \qquad\quad + \frac{10}{189}(M^{(4)}_{{j\rangle}bc} S^{(2)}_{ab}-M^{(3)}_{{j\rangle}bc} S^{(3)}_{ab}) \biggr]+ \frac{\ud}{\ud u}\biggr[ -\frac{25}{1134}M^{(2)}_{ab} M^{(4)}_{ijab} +\frac{20}{189}\epsilon_{ac{\langle i}}M^{(3)}_{{j\rangle}bc} S^{(2)}_{ab} \nonumber\\ 
    &\qquad\quad -\frac{5}{42}\epsilon_{ac{\langle i}}S^{(3)}_{{j\rangle}bc} M^{(2)}_{ab} + \frac{144}{63} S^{(2)}_{a{\langle i}} S^{(2)}_{{j\rangle}a} +\frac{25}{324} M^{(3)}_{ab{\langle i}}M^{(3)}_{{j\rangle}ab} \biggr] +\frac{\ud^2}{\ud u^2}\biggr[ -\frac{1}{432}M_{ab} M^{(5)}_{ijab} \nonumber\\
    &\qquad\quad + \frac{3}{432}M^{(1)}_{ab} M^{(4)}_{ijab}+\frac{5}{432}M^{(2)}_{ab} M^{(3)}_{ijab}+ \frac{1}{144}M^{(3)}_{ab} M^{(2)}_{ijab}+ \frac{235}{378}M^{(4)}_{ab} M^{(1)}_{ijab} \nonumber\\
    & \quad\qquad+ \frac{91}{216}M^{(5)}_{ab} M_{ijab} + \epsilon_{ac{\langle i}}\biggr( \frac{1}{189}M^{(4)}_{{j\rangle}bc} S_{ab} -\frac{3}{189}M^{(3)}_{{j\rangle}bc} S^{(1)}_{ab} -\frac{5}{63}M^{(2)}_{{j\rangle}bc} S^{(2)}_{ab}  \nonumber \\ 
    &\qquad\quad + \frac{55}{189}M^{(1)}_{{j\rangle}bc} S^{(3)}_{ab} - \frac{10}{63}M_{{j\rangle}bc} S^{(4)}_{ab} + \frac{1}{168}S^{(4)}_{{j\rangle}bc} M_{ab} +\frac{5}{168}S^{(3)}_{{j\rangle}bc} M^{(1)}_{ab} \nonumber \\ 
    &\qquad\quad+ \frac{25}{168}S^{(2)}_{{j\rangle}bc} M^{(2)}_{ab} +\frac{73}{168}S^{(2)}_{{j\rangle}bc} M^{(3)}_{ab} +\frac{65}{84}S_{{j\rangle}bc} M^{(4)}_{ab}   \biggr)  + \frac{80}{63}S_{a{\langle i}} S^{(3)}_{{j\rangle}a} \nonumber\\
    &\qquad\quad+\frac{16}{63}S^{(1)}_{a{\langle i}} S^{(2)}_{{j\rangle}a} +\frac{144}{63}S^{(3)}_{a{\langle i}} S_{{j\rangle}a}  - \frac{5}{252}M_{ab{\langle i}}M^{(5)}_{{j\rangle}ab}  +\frac{25}{378}M^{(1)}_{ab{\langle i}}M^{(4)}_{{j\rangle}ab} \nonumber\\
    &\qquad\quad \left. -\frac{55}{756}M^{(2)}_{ab{\langle i}}M^{(3)}_{{j\rangle}ab} + \frac{5}{42}S_a S^{(3)}_{ija}  \biggr]\right\}, 
\end{align}
and for the mass octupole moment the result is
\begin{align} \label{eqn:octupole PN exp decomposed}
    \mathcal{U}_{ijk} (U) =& M^{(3)}_{ijk} (U) + 2M \left(\frac{G}{c^3} \right)
    \int^{U}_{-\infty} \ud \tau\biggr[ \ln
    \left(\frac{c(U-\tau)}{2r_0}\right)+\frac{97}{60} \biggr]
    M^{(5)}_{ijk} (\tau)\nonumber \\ 
    & + \left(\frac{G}{c^5}\right) \biggr\{
    \int^{U}_{-\infty} \ud \tau \biggr[-\frac{1}{3}M^{(3)}_{a\langle
    i} M^{(4)}_{jk\rangle a} -\frac{4}{5}\epsilon_{ab\langle i}
    M^{(3)}_{ja} S^{(3)}_{k\rangle b}
    \biggr](\tau)\nonumber\\ 
    & \qquad\quad+\frac{1}{12}(M^{(2)}_{a\langle
    i}M^{(4)}_{jk\rangle a}-M^{(3)}_{a\langle i}M^{(3)}_{jk\rangle
    a})+ \frac{11}{12}\frac{\ud}{\ud u}\biggr(M^{(2)}_{a\langle i}M^{(3)}_{jk\rangle
    a}\biggr)\nonumber\\ 
    & \qquad\quad+ \frac{\ud^2}{\ud u^2}\biggr(-\frac{1}{4} M^{(2)}_{a\langle i}M^{(2)}_{jk\rangle
    a} -\frac{11}{12}M^{(3)}_{a\langle i}M^{(1)}_{jk\rangle
    a} +\frac{1}{12}M^{(4)}_{a\langle i}M_{jk\rangle
    a}\nonumber\\ 
    & \qquad\quad+ \frac{1}{4}M_{a\langle i}M^{(4)}_{jk\rangle
    a} +\frac{12}{5}S_{\langle
  i}S^{(2)}_{jk\rangle} \biggr)+ \frac{1}{5}\epsilon_{ab\langle
    i}\biggr[18 \frac{\ud}{\ud u}\biggr(M^{(2)}_{ja}S^{(2)}_{k\rangle b}\biggr) \nonumber\\
    & \qquad\quad+\frac{\ud^2}{\ud u^2}\biggr(-9M^{(1)}_{ja}S^{(2)}_{k\rangle b}+ M^{(2)}_{ja}S^{(1)}_{k\rangle b}+ M^{(3)}_{ja}S_{k\rangle b}\nonumber\\ 
    & \qquad\quad-9 M_{ja}S^{(3)}_{k\rangle b}
    -\frac{9}{4}S_{a}M^{(3)}_{jk\rangle b}\biggr)\biggr]\biggr\}+
    \mathcal{O}\biggr(\frac{1}{c^6}\biggr) .
\end{align}

\section{Radiative modes for compact binaries in quasi-circular orbits} \label{app:radModesBinaries}

Here, we list the radiative modes that were used to calculate the leading PN order of the moments of the news and the corresponding shear. 
These expressions were obtained from Sec.~3.4.5 of~\cite{Blanchet:2013haa}.
For the majority of our calculations, we used the leading, ``Newtonian'' expressions for these modes given below:
\begin{subequations}
    \begin{align}
        & U_{20} = \frac{4}{7}\sqrt{\frac{5\pi}{3}}M \eta x , \qquad 
        V_{21}^\mathrm{N} =  \frac{8}{3} \sqrt{\frac{2 \pi }{5}} \delta m \eta  x^{3/2} e^{-i\phi} ,  \qquad 
        U_{22}^\mathrm{N} = -8\sqrt{\frac{2\pi}{5}}M\eta x e^{-i2\phi} , \\
        & U_{31}^\mathrm{N} = -\frac{2}{3} i \sqrt{\frac{\pi }{35}} \delta m  \eta  x^{3/2} e^{-i\phi},  \qquad
        U_{33}^\mathrm{N} = 6i \sqrt{\frac{3\pi}{7}} \delta m \eta x^{3/2}e^{-i3\phi}, \\
        & V_{30} =  \frac{32}{5} \sqrt{\frac{3 \pi }{35}} M\eta ^2 x^{7/2} , \qquad 
        V_{32}^\mathrm{N} = -\frac{8}{3}  i M \sqrt{\frac{2 \pi }{7}} (1-3 \eta ) \eta x^2 e^{-i2\phi} , \\
        & U_{40} =  -\frac{1}{63} \sqrt{\frac{\pi }{5}} M \eta x \, \qquad
        U_{42}^\mathrm{N} = -\frac{8}{63} \sqrt{2 \pi } \eta (1-3\eta ) M x^2 e^{-i2\phi} .
    \end{align}
\end{subequations}
We left the ``N'' label off the $m=0$ modes, because these modes are computed from the zeroth or first fluxes, which go beyond linearized gravity.

However, to compute the entries in Table~\ref{tab: Coefficients of Shear contribution } given by Eqs.~\eqref{eqn: COM U20} and~\eqref{eqn: Grad* U30}, we needed to go to higher PN orders, because the leading terms cancel.
There are subleading terms that that are out of phase by $\pi/2$ from the leading terms that do not cancel, however.
These terms are related to radiation reaction effects, and the ones that we require appear at 1.5~PN order relative to the ``Newtonian'' mode.
We give these terms the superscript ``RR'' for radiation reaction, and we list them below: 
\begin{subequations}
 \begin{align}
     & V_{21}^\mathrm{RR} = -\frac{i}{2} V_{21}^\mathrm{N} (1+\ln 16)x^{3/2} , \qquad 
     U_{31}^\mathrm{RR} = -i U_{31}^N \left(\frac 75 + \ln 4 \right) x^{3/2} , \\
     & V_{32}^\mathrm{RR} = -\frac{i}{1-3\eta} V_{32}^\mathrm{N} \biggr(3-\frac{66 \eta }{5}\biggr) x^{3/2} , \qquad 
     U_{42}^\mathrm{RR}=  -\frac{i}{1-3\eta} U_{42}^\mathrm{N} \biggr(\frac{21-84\eta}{5}\biggr) x^{3/2} .
 \end{align}
\end{subequations}

\section{Relation between definitions of the moments of the news} \label{app:MomentRelations}

The curve deviation observable in~\cite{Grant:2021hga} was written in terms of the following moments:
\begin{equation} \label{eqn: Mellin moment}
    \ord{n}{\mathcal N}_{AB} ^{\mathcal M} (u_1, u_0)= \frac{1}{n !}\int _{u_0} ^{u_1} \ud u_2 (u_2 -u_0)^n N_{AB}(u_2). 
\end{equation}
It was noted in~\cite{Compere:2022zdz,Grant:2023jhd} that these moments are related to the Mellin transform of the news tensor.
We made a remark in Sec.~\ref{sec:Moments of the news} that the definition of moments in this paper differs from those in Eq.~\eqref{eqn: Mellin moment}. 
However, the so-called ``Mellin moments'' in Eq.~\eqref{eqn: Mellin moment} can be written in terms of the moments defined in Eq.~\eqref{eqn:Mom news} by using Cauchy's formula for multiple integration.
After a short calculation, one can show that the relation is
\begin{align}
     \ord{n}{\mathcal N}_{AB} ^{\mathcal M} (u_1, u_0)= \sum _{m=0}^n \frac{(-1)^m}{(n-m)!} (u_1-u_0)^{n-m} \ord{m}{\mathcal N}_{AB} (u_1, u_0).
\end{align}
Moreover, by Cauchy's formula for multiple integration, the moments of the news in this paper are the same as the ``Cauchy moments'' which appear in~\cite{Grant:2023jhd}, up to minor differences in the exact definition of ``news''.

\section{Tensors on the 2-sphere} \label{app:STFtensorIdentities}

In this appendix, we provide several properties and relationships satisfied by tensors on the 2-sphere, including a number of results for tensor and spin-weighted spherical harmonics.
We used many of these results throughout this paper.

Tensors on the 2-sphere can be written on a null, complex dyad $\{m^A, \bar m^A\}$, which satisfies the conditions
\begin{equation}
    h_{AB} m^A {m}^B = 0, \qquad h_{AB} m^A \bar m^B = 1.
\end{equation}
In terms of this dyad, the metric on the 2-sphere can be written as
\begin{equation}
    h_{AB} = 2 m_{(A} \bar{m}_{B)}.
\end{equation}
Here, we focus on STF tensors, as  many of the tensors which we consider in this paper are STF. Any STF tensor $S_{A_1 \cdots A_s}$ on the 2-sphere has only 2 degrees of freedom, since it can be shown to be written as
\begin{equation}
    S_{A_1 \cdots A_s} = \bar S m_{A_1} \cdots m_{A_s} + S \bar m_{A_1} \cdots \bar m_{A_s},
\end{equation}
where
\begin{equation} \label{eqn:m_contraction}
    S \equiv m^{A_1} \cdots m^{A_s} S_{A_1 \cdots A_s}.
\end{equation}

In general, for some tensor field $S_{A_1 \cdots A_p}$, if it transforms as  $S_{A_1 \cdots A_p} \to e^{is\theta} S_{A_1 \cdots A_p}$ when $m^A \to e^{i\theta} m^A$, then the tensor field is said to have \emph{spin weight} $s$.
While discussions of spin weight (e.g.,~\cite{Geroch:1973am}) often are restricted to scalar quantities, we find it convenient to also consider tensors of spin weight, which have been discussed elsewhere: for example, at the beginning of Sec.~4.12 of~\cite{penrose1987spinors}.
Examples of spin-weighted tensors include $m^A$ itself (spin weight $1$), as well as a tensor that has only a subset of its indices contracted with dyad vectors, such as $\bar m^A C_{AB}$ (spin weight $-1$).
If $S_{A_1 \cdots A_s}$ has zero spin weight, then the scalar in Eq.~\eqref{eqn:m_contraction} has spin weight $s$.

A convenient basis for STF tensors on the 2-sphere are the \emph{tensor harmonics}, which are defined by
\begin{subequations}
   \begin{align}
       (T^{\mathrm E}_{lm})_{A_1 \cdots A_s}&= 2^{(s-1)/2}\sqrt{\frac{(l-s)!}{(l+s)!}} \STF(\mathscr{D}_{A_1} \cdots \mathscr{D}_{A_s} Y_{lm}), \\
       (T^{\mathrm B}_{lm})_{A_1 \cdots A_s}&= 2^{(s-1)/2}\sqrt{\frac{(l-s)!}{(l+s)!}} \STF(\epsilon_{BA_1} \mathscr{D}_{A_2} \cdots \mathscr{D}_{A_s}\mathscr{D}^{B} Y_{lm}).
   \end{align}
\end{subequations}
On contracting the tensor harmonics with members of the dyad, one can show that the tensor harmonics can be written in terms of spin-weighted spherical harmonics:
\begin{subequations}{\label{eqn: Tensor har exp}}
    \begin{align}
     (T^{\mathrm E}_{lm})_{A_1 \cdots A_s}&=\frac{1}{\sqrt{2}}[{_{-s}Y_{lm}}m_{A_1} \cdots m_{A_s} + (-1)^s {_{s}Y_{lm}}\bar{m}_{A_1} \cdots \bar{m}_{A_s}], \\
     (T^{\mathrm B}_{lm})_{A_1 \cdots A_s}&=-i\frac{1}{\sqrt{2}}[{_{-s}Y_{lm}}m_{A_1} \cdots m_{A_s} - (-1)^s {_{s}Y_{lm}}\bar{m}_{A_1} \cdots \bar{m}_{A_s}],
\end{align}
\end{subequations}
where the spin-weighted spherical harmonics ${}_s Y_{lm}$ are defined by
\begin{equation} \label{eqn:swsh_def}
    {_{s}Y_{lm}}= 2^{|s|/2} \sqrt{\frac{(l - |s|)!}{(l + |s|)!}} 
    \begin{cases}
        (-\eth)^s Y_{lm} & s \geq 0 \\
        (\bar{\eth})^s Y_{lm} & s \leq 0
    \end{cases},
\end{equation}
and where the derivative operator $\eth$ is defined, when acting on a spin weight $s$ tensor $S_{A_1 \cdots A_p}$, by
\begin{equation} \label{eq:raiseTensor}
    \eth S_{A_1 \cdots A_p} \equiv [m^B \mathscr D_B - s \bar m^C m^B \mathscr D_B (m_C)] S_{A_1 \cdots A_p}.
\end{equation}
It follows that $\eth$ raises the spin-weight of any tensor on which it acts, $\bar \eth$ lowers the spin-weight, and
\begin{equation}
    \eth m^A = 0, \qquad \bar \eth m^A = 0.
\end{equation}
For example, using Eq.~\eqref{eqn:swsh_def}, one can derive the following formula for raising and lowering the spin-weighted spherical harmonics:
\begin{equation} \label{eq:raise_sYlm}
    \eth {}_s Y_{lm} = -\sqrt{\frac{(l - s)(l + s + 1)}{2}} {}_{s + 1} Y_{lm}, \quad \bar \eth {}_s Y_{lm} = \sqrt{\frac{(l + s)(l - s + 1)}{2}} {}_{s - 1} Y_{lm}.
\end{equation}

We now discuss the raising and lowering the tensor rank of tensor harmonics with the covariant derivative operator $\mathscr D_A$.
First, directly from the harmonics' definitions, we have that raising is given by
\begin{gather} \label{eq:raise_T_EB}
    \STF[\mathscr{D}_{A_1}(T^{\mathrm I}_{lm})_{A_2 \cdots A_{s+1}}] = \sqrt{\frac{(l-s)(l+s+1)}{2}} (T^{\mathrm I}_{lm})_{A_1 \cdots A_{s+1}},\\
    (T^{\mathrm E}_{lm})_A = \frac{1}{\sqrt{l(l+1)}} \mathscr D_A Y_{lm}, \qquad (T^{\mathrm B}_{lm})_A = \frac{1}{\sqrt{l(l+1)}} \epsilon_{BA} \mathscr D^B Y_{lm},
\end{gather}
where $\mathrm I = \mathrm E, \mathrm B$.
Deriving the lowering relationship takes a few more steps.
Writing $\mathscr{D}^{B}(T^{\mathrm I}_{lm})_{B A_1 \cdots A_{s-1}} = 2m^{(B}\bar m^{C)} \mathscr D_B (T^{\mathrm I}_{lm})_{C A_1 \cdots A_{s-1}}$, then using Eq.~\eqref{eqn: Tensor har exp}, the product rule,  Eq.~\eqref{eq:raiseTensor} (and its complex conjugate) and Eq.~\eqref{eq:raise_sYlm}, allows us to derive the following expressions for lowering tensor harmonics:
\begin{subequations}\label{eq:lower_T_EB}
\begin{align}
    \mathscr{D}^{B}(T^{\mathrm I}_{lm})_{B A_1 \cdots A_{s-1}} &= -\sqrt{\frac{(l+s)(l-s+1)}{2}}(T^{\mathrm I}_{lm})_{A_1 \cdots A_{s-1}}, \qquad (s > 1) \\
    \mathscr{D}^A (T^{\mathrm E}_{lm})_A &= \epsilon_{AB} \mathscr{D}^A (T^{\mathrm B}_{lm})^B = -\sqrt{l(l+1)}Y_{lm}, \\
    \epsilon_{AB}\mathscr{D}^A (T^{\mathrm E}_{lm})^B &= \mathscr{D}^A (T^{\mathrm B}_{lm})_A = 0.
\end{align}
\end{subequations}
Finally, the action of the Laplacian on the tensor harmonics is given by
\begin{equation} \label{eq:D2Tlm}
    \mathscr D^2 (T^{\mathrm I}_{lm})_{A_1 \cdots A_s} = [s^2 - l(l + 1)] (T^{\mathrm I}_{lm})_{A_1 \cdots A_s},
\end{equation}
which can be obtained through a calculation similar to that described for the lowering relationship.
The eigenvalue of $\mathscr D^2$ on the right-hand side of Eq.~\eqref{eq:D2Tlm} reduces to the usual factor of $-l(l + 1)$ when restricting to the scalar harmonics.

We now derive expressions for the contraction of two tensor harmonics of equal rank in terms of a sum of scalar harmonics.
Using Eq.~\eqref{eqn: Tensor har exp}, we obtain
\begin{equation} \label{eq:2Tcontracted}
    (T^{\mathrm I'}_{l'm'})_{A_1 \cdots A_s} (T^{\mathrm I''}_{l''m''})^{A_1 \cdots A_S} = \frac{(-1)^s}{2} \left(c^{\mathrm I' \mathrm I''} {}_{-s} Y_{l'm'}\; {}_s Y_{l''m''} + \overline{c^{\mathrm I' \mathrm I''}} {}_s Y_{l'm'}\; {}_{-s} Y_{l''m''}\right),
\end{equation}
where
\begin{equation}
    c^{\mathrm I \mathrm I'} = \begin{cases}
        1 & \mathrm I = \mathrm I' = \mathrm E, \mathrm B \\
        i & \mathrm I = \mathrm E, \ \mathrm I' = \mathrm B \\
        -i & \mathrm I = \mathrm B, \ \mathrm I' = \mathrm E
    \end{cases}.
\end{equation}
Next, we can use the fact that
\begin{equation} \label{eq:sYlmProduct}
    {}_{-s} Y_{l'm'}\; {}_s Y_{l''m''} = \sum_{l, m} \mathscr C^{lms}_{l'm'l''m''} Y_{lm},
\end{equation}
for the coefficient $\mathscr C^{lms}_{l'm'l''m''}$ defined in Eq.~\eqref{eqn:Def Coeff C}, which has the property that
\begin{equation}
    \mathscr C^{lms}_{l'm'l''m''} = (-1)^{l + l' + l''} \mathscr C^{lm(-s)}_{l'm'l''m''}.
\end{equation}
This allows us to write Eq.~\eqref{eq:2Tcontracted} as
\begin{equation}
    (T^{\mathrm I'}_{l'm'})_{A_1 \cdots A_s} (T^{\mathrm I''}_{l''m''})^{A_1 \cdots A_s} = \frac{(-1)^s}{2} \sum_{l, m} \eta^{\mathrm E \mathrm I' \mathrm I''}_{ll'l''} \mathscr C^{lms}_{l'm'l''m''} Y_{lm},
\end{equation}
where
\begin{equation}
    \eta^{\mathrm E \mathrm I' \mathrm I''}_{ll'l''} = c^{\mathrm I' \mathrm I''} + (-1)^{l + l' + l''} \overline{c^{\mathrm I' \mathrm I''}}.
\end{equation}
The values for $\eta^{\mathrm E \mathrm I' \mathrm I''}_{ll'l''}$ are given in the first three cases of Eq.~\eqref{eq:etaDef}.

The tensor harmonics satisfy the following property on contracting with the two-dimensional Levi-Civita tensor:
\begin{align}\label{eqn: Conj TH property}
    \epsilon^{B}{}_{A_1} (T^{\mathrm I'})_{B A_2 \cdots A_n} = \epsilon^{\mathrm I' \mathrm I''} (T^{\mathrm I''})_{A_1 \cdots A_n},
\end{align}
where we define
\begin{align}
    \epsilon^{\mathrm I' \mathrm I''} = \begin{cases}
        1 & \mathrm I' = \mathrm E, \ \mathrm I'' = \mathrm B \\
        -1 & \mathrm I' = \mathrm B, \ \mathrm I'' = \mathrm E \\
        0 & \textrm{otherwise}
    \end{cases}.
\end{align}
Also, our definition of $\eta$ has the following property:
\begin{align}\label{eqn: eta property}
    \epsilon^{\tilde{\mathrm I} \mathrm I''} \eta^{\mathrm I \mathrm I' \tilde{\mathrm I}}_{ll'l''} = \epsilon^{\bar{\mathrm I} \mathrm I} \eta^{\bar{\mathrm I} \mathrm I' \mathrm I''}_{ll'l''}.
\end{align}
Thus, using Eqs.~\eqref{eqn: Conj TH property} and~\eqref{eqn: eta property}, one can evaluate the following contraction which was used for computing the magnetic parts of the fluxes:
\begin{align}
    \epsilon_B{}^{A_1} (T^{\mathrm I'}_{l'm'})_{A_1 \cdots A_s} (T^{\mathrm I''}_{l''m''})^{B A_2 \cdots A_s} = \frac{(-1)^s}{2} \sum_{l, m} \eta^{\mathrm B \mathrm I' \mathrm I''}_{ll'l''} \mathscr C^{lms}_{l'm'l''m''} Y_{lm}.
\end{align}
The values for $\eta^{\mathrm B \mathrm I' \mathrm I''}_{ll'l''}$ are given in the last three cases of Eq.~\eqref{eq:etaDef}.

\end{document}